%% file: main.tex
\documentclass{article}

\usepackage[final]{neurips_2025_ml4ps}

\usepackage[utf8]{inputenc} 
\usepackage[T1]{fontenc}    
\usepackage{hyperref}       
\usepackage{url}            
\usepackage{booktabs}       
\usepackage{amsfonts}       
\usepackage{nicefrac}       
\usepackage{microtype}      
\usepackage{xcolor}         

\usepackage{lineno}
\usepackage{graphicx}
\usepackage{amsmath}
\usepackage{booktabs}
\usepackage{caption}
\usepackage{hyperref}
\usepackage{multirow}
\usepackage{subcaption}
\usepackage{natbib}
\usepackage{enumitem}
\usepackage{makecell} 
\usepackage{arydshln}

\usepackage[most]{tcolorbox}
\usepackage{etoolbox} 
\tcbset{
  colback=white,
  colframe=black!40,
  boxrule=0.5pt,
  arc=2mm
}
\newtcblisting{promptlisting}{
  breakable,
  listing only,
  listing options={
    basicstyle=\ttfamily\scriptsize,
    breaklines=true,
    showstringspaces=false,
    literate={
      {•}{{\textbullet}}1
      {’}{{\textquotesingle}}1
      {–}{{--}}1
      {—}{{---}}1
      {“}{{\textquotedblleft}}1
      {”}{{\textquotedblright}}1
    }
  }
}

\title{Radio Astronomy in the Era of Vision-Language Models: Prompt Sensitivity and Adaptation}

\author{%
  Mariia Drozdova \\
  University of Geneva, Switzerland \\
  \texttt{mariia.drozdova@unige.ch} 
  \And
  Erica Lastufka \\
  University of Geneva, Switzerland \\
  \texttt{erica.lastufka@unige.ch}
  \And
  Vitaliy Kinakh \\
  University of Geneva, Switzerland \\
  \texttt{vitaliy.kinakh@unige.ch}
  \And
  Taras Holotyak \\
  University of Geneva, Switzerland \\
  \texttt{taras.holotyak@unige.ch}
  \And
  Daniel Schaerer \\
  University of Geneva, Switzerland \\
  \texttt{daniel.schaerer@unige.ch}
  \And
  Slava Voloshynovskiy \\
  University of Geneva, Switzerland \\
  \texttt{svolos@unige.ch}
}

\begin{document}

\maketitle

\begin{abstract}
Vision–Language Models (VLMs), such as recent Qwen and Gemini models, are positioned as general-purpose AI systems capable of reasoning across domains. Yet their capabilities in scientific imaging, especially on unfamiliar and potentially previously unseen data distributions, remain poorly understood. In this work, we assess whether generic VLMs, presumed to lack exposure to astronomical corpora, can perform morphology-based classification of radio galaxies using the MiraBest \mbox{FR-I/FR-II} dataset. We explore prompting strategies using natural language and schematic diagrams, and, to the best of our knowledge, we are the first to introduce visual in-context examples within prompts in astronomy. Additionally, we evaluate lightweight supervised adaptation via LoRA fine-tuning. Our findings reveal three trends: (i) even prompt-based approaches can achieve good performance, suggesting that VLMs encode useful priors for unfamiliar scientific domains; (ii) however, outputs are highly unstable, i.e. varying sharply with superficial prompt changes such as layout, ordering, or decoding temperature, even when semantic content is held constant; and (iii) with just 15M trainable parameters and no astronomy-specific pretraining, fine-tuned Qwen-VL achieves near state-of-the-art performance (3\% Error rate), rivaling domain-specific models. These results suggest that the apparent “reasoning” of VLMs often reflects prompt sensitivity rather than genuine inference, raising caution for their use in scientific domains. At the same time, with minimal adaptation, generic VLMs can rival specialized models, offering a promising but fragile tool for scientific discovery.
\end{abstract}

\section{Introduction}
\label{sec:introduction}
Vision–language models (VLMs), such as recent Gemini~\cite{comanici2025gemini}, Qwen-VL~\cite{team2024qwen2}, and GPT-4o~\cite{openai_chatgpt_2023}, show strong performance on general multimodal tasks, but their utility in scientific settings, especially when applied directly to domain-specific imagery, remains an open question. In astronomy, vision foundation models (VFMs) pretrained on natural images show promise when fine-tuned on scientific data~\cite{lastufka2024vision, drozdova2024semi}. Even stronger results come from domain-specific pretraining, such as the vision-only model~\cite{slijepcevic2024radio}(which we refer to as AstroVFM), the multimodal AstroM3~\cite{rizhko2025astrom3} and CosmoCLIP~\cite{imam2024cosmoclip}. This raises the question: \textit{Can general-purpose VLMs adapt to scientific image classification with little or no supervision?}

VLMs have been explored in astronomy by~\cite{zaman2025astrollava}, which focuses on optical image captioning and visual question answering. We instead focus on binary morphology classification using the MiraBest dataset~\cite{porter2023mirabest}: radio galaxies classified into Fanaroff–Riley types (FR-I vs FR-II). Riggi et al.~\cite{riggi2025evaluating} evaluated VLMs on MiraBest, reporting low performance (e.g., $\sim$30\% F1 for Qwen2-VL-2B, $\sim$20\% for 7B) using base models in a chat interface. Their fine-tuned LLaVA model also remained below 30\%. In contrast, we use the instruction-tuned Qwen2-VL-7B-Instruct and show that prompt design alone yields 84\% Macro-F1, and LoRA tuning pushes it to 97\%.

We explore several prompting strategies : natural language descriptions, schematic diagrams, and, to the best of our knowledge for the first time in this domain, retrieval-augmented prompting. Inspired by the success of retrieval-augmented generation (RAG) in LLMs~\cite{lewis2020retrieval} and recent theory showing that in-context examples can act like implicit weight updates~\cite{dherin2025learning}, we retrieve support images in CLIP space~\cite{radford2021learning} and embed them in the prompt. In zero-shot settings (see also~\cite{tanoglidis2024first}), Gemini achieves errors as low as 14\%. Open models perform worse without examples, but improve notably when conditioned on retrieved ones. However, results reveal high sensitivity to prompt layout, decoding temperature, and example ordering, as presented in Sec. \ref{sec:results}.


Finally, we fine-tune Qwen2-VL-7B-Instruct using LoRA~\cite{hu2022lora}. On the full training set, this reduces the error to 3\%, which is close to the 2\% error achieved by a domain-pretrained ResNet (AstroVFM)\cite{slijepcevic2024radio}, self-superised through BYOL\cite{grill2020bootstrap} on the RGZ DR1 dataset~\cite{wong_2024_14195049} and also fine-tuned on MiraBest.

Our results suggest that generalist VLMs encode useful representations for radio astronomy imagery, but their success depends critically on the prompt construction and adaptation method. 

\section{Methods}
\label{sec:methods}

\paragraph{Prompting Setup.}
We evaluate several VLMs on the task of classifying radio galaxies into Fanaroff–Riley types I and II (details in Appendix \ref{app:dataset}). Each test query ends with a radio image and a prompt requesting classification as FR-I or FR-II. The context provided before the image varies across five prompting strategies: 
\textbf{(1) Text}, natural language descriptions of each class (zero-shot); 
\textbf{(2) Diagram}, the same text augmented with an abstract schematic (Fig.~\ref{fig:fr_schematic}); 
\textbf{(3) Fixed-Imgs}, four labeled support images, identical across test queries; 
\textbf{(4) kNN-Imgs}, five labeled nearest neighbors retrieved per test sample in CLIP space~\cite{radford2021learning}; and 
\textbf{(5) kNN-Balanced}, a balanced version of (4), retrieving labeled neighbors equally from both classes, a setup not commonly explored in prior VLM work. In strategies (3)-(5), both support images and their labels are included in the prompt. We evaluate each setup in two variants: chain-of-thought (CoT), prompting models to explain their reasoning, and noCoT, requesting only a direct label. To probe decoding variability, we vary temperature: lower values yield deterministic outputs, while higher ones increase sampling diversity, potentially revealing reasoning instability.

\paragraph{Fine-Tuning.}
We explore whether lightweight adaptation improves VLM performance. We fine-tune Qwen2-VL-7B-Instruct using LoRA~\cite{hu2022lora}, updating $\sim$15M parameters across vision and language modules. LoRA hyperparameters ($r=16$, $\alpha=64$, dropout=0.3) are chosen via grid search. Training runs for 100 epochs, with test metrics reported at the first minimum of training loss. We compare results across varying training set sizes to a supervised ResNet trained from scratch and AstroVFM fine-tuned, both using \texttt{MiraBest-confident} as in~\cite{slijepcevic2024radio}. Our model is fine-tuned on the same dataset.
\section{Results}
\label{sec:results}
\begin{figure}[t]
\centering
\includegraphics[width=0.99\linewidth]{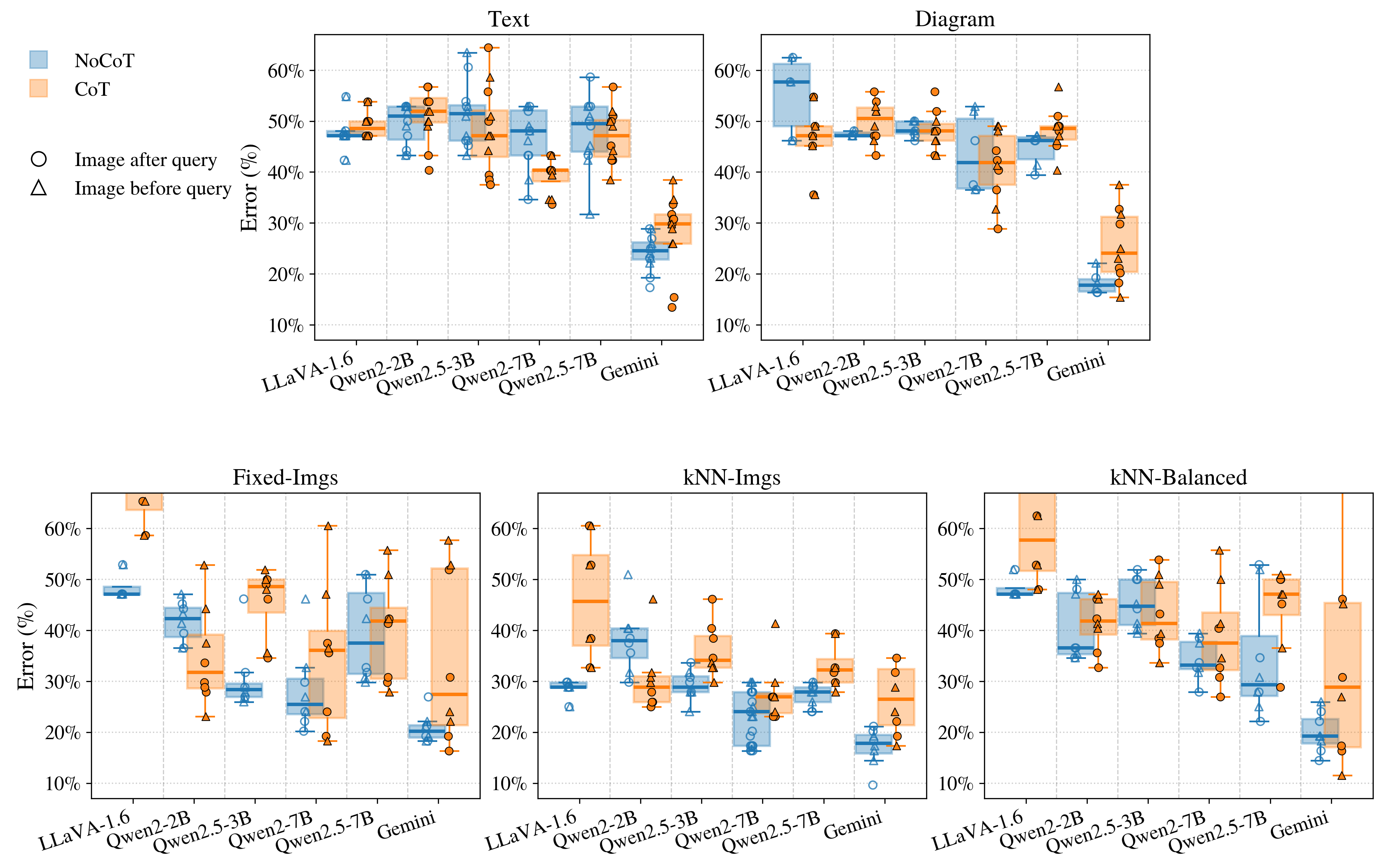}
\caption{
Test error rates across prompting strategies with/without CoT. Boxplots summarize variation across prompts and image placement with respect to the query question. More details in Appendix\ref{app:analysis}.
}

\label{fig:prompting_boxplots}
\end{figure}

\begin{figure}[t]
    \centering
    \begin{subfigure}[b]{0.32\linewidth}
        \centering
        \includegraphics[height=3.6cm]{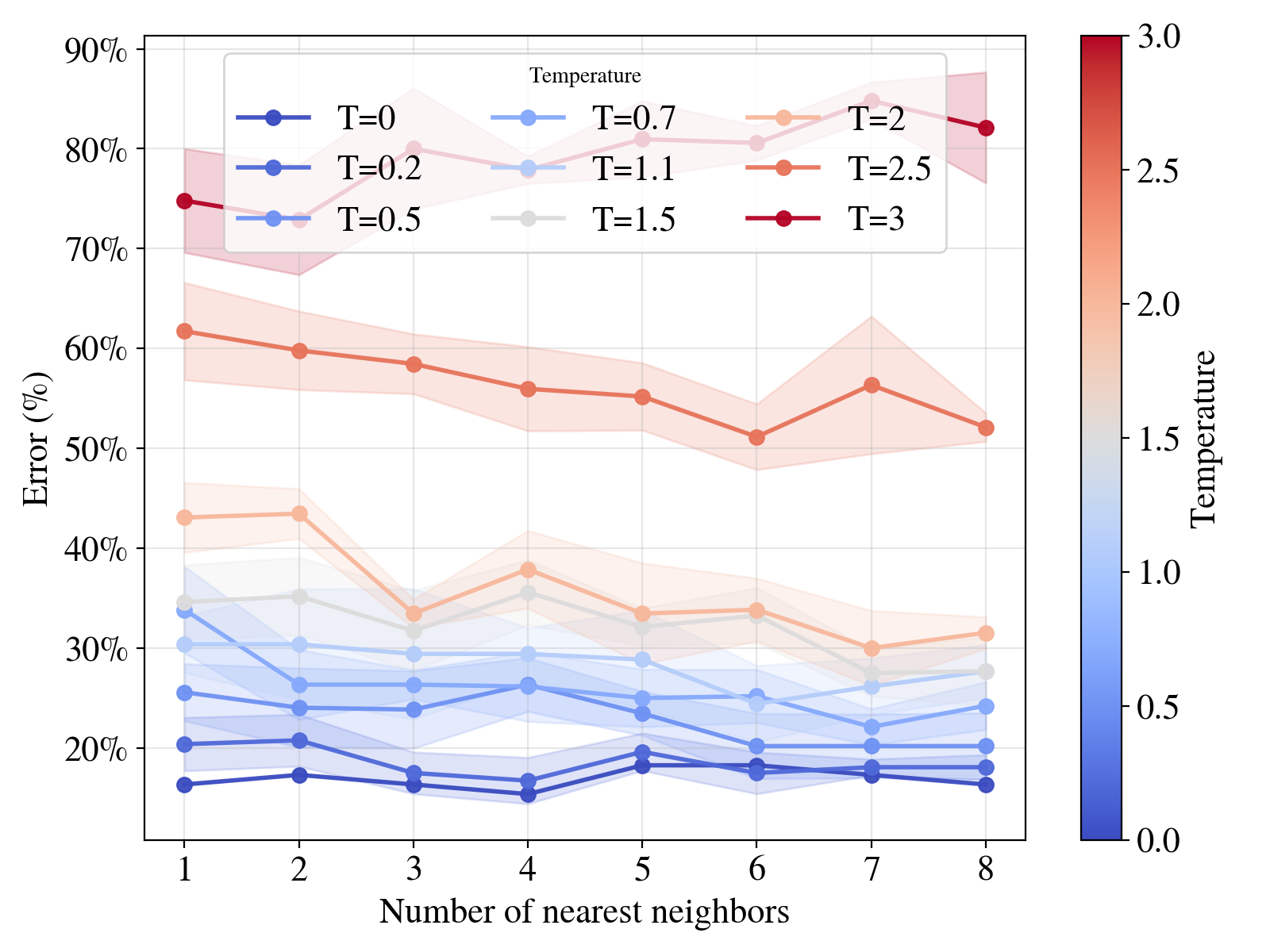}
        \caption{}
        \label{fig:temperature_influence}
    \end{subfigure}
    \hfill
    \begin{subfigure}[b]{0.32\linewidth}
        \centering
        \includegraphics[height=3.6cm]{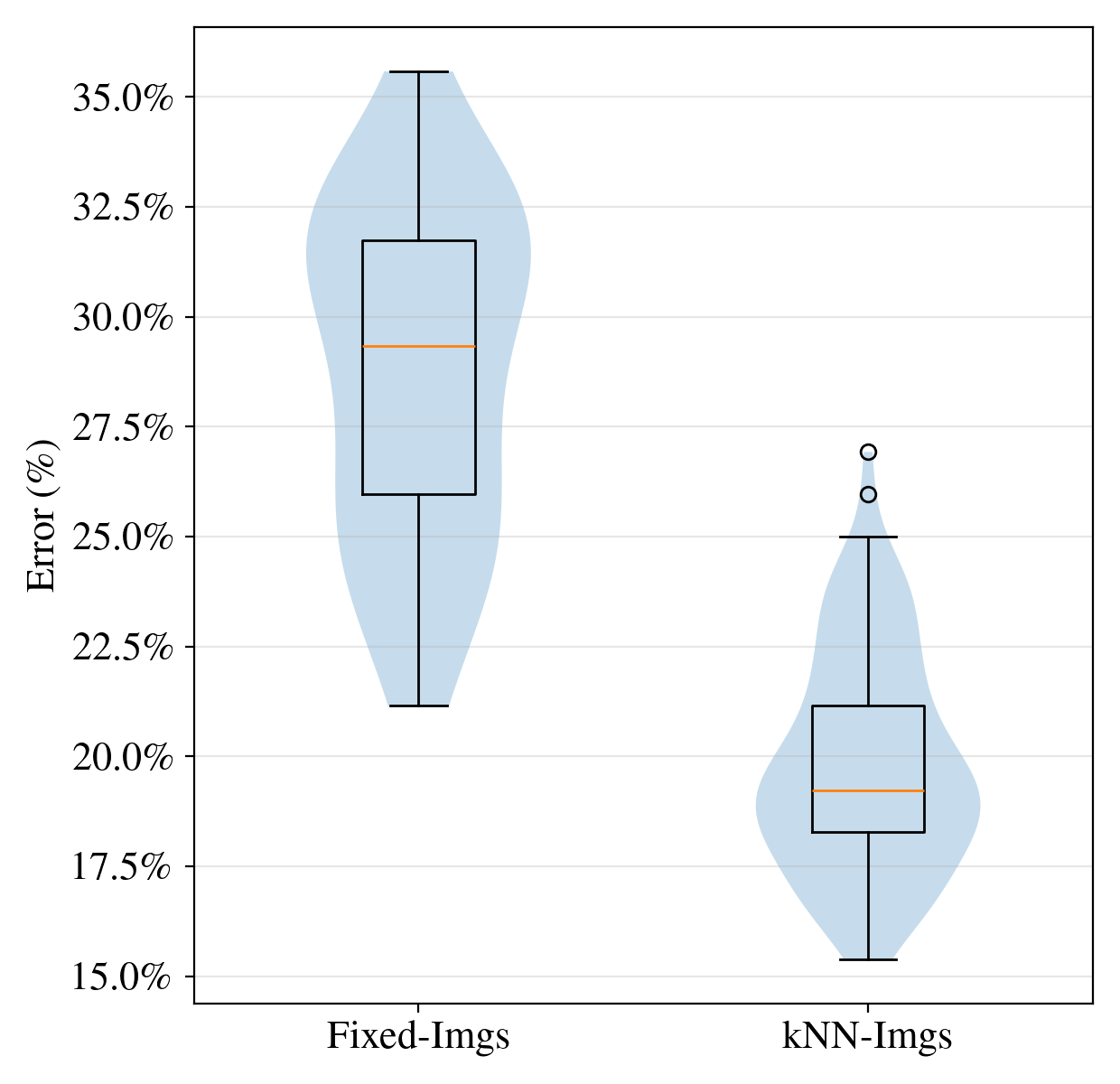}
        \caption{}
        \label{fig:permutations_instability}
    \end{subfigure}
    \hfill
    \begin{subfigure}[b]{0.32\linewidth}
        \centering
        \includegraphics[height=3.6cm]{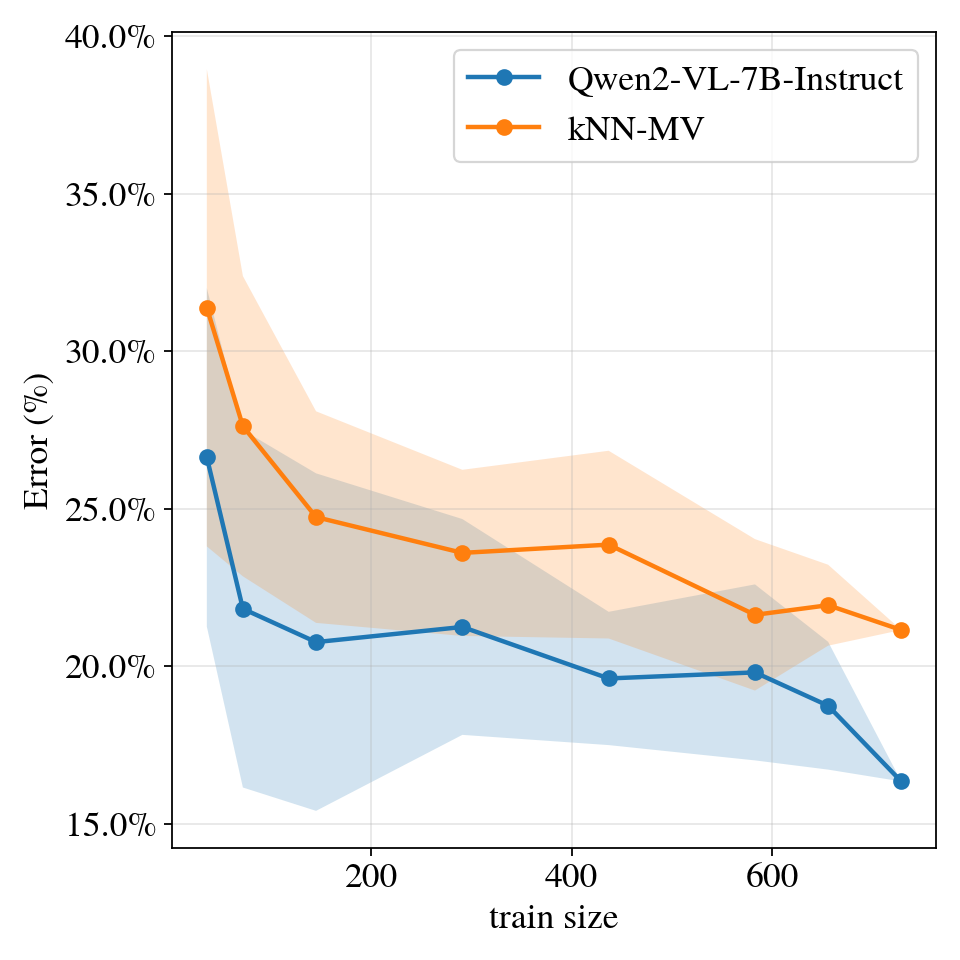}
        \caption{}
        \label{fig:knn_comparison}
    \end{subfigure}
\caption{
Stability of example-conditioned prompts: 
(a) Error vs. number of retrieved neighbors; lower temperatures reduce error and spread;  
(b) Error across all permutations: \texttt{Fixed-Imgs} is less stable than \texttt{kNN-Imgs}; 
(c) \texttt{kNN-Imgs} outperform \texttt{kNN} majority voting by ~5 points across train sizes.
}
    \label{fig:stability}
\end{figure}

\paragraph{Prompting}

Fig.~\ref{fig:prompting_boxplots} summarizes test error across prompting types. In Text and Diagram settings, Gemini performs best, reaching $\sim$14\% error. Open models lag behind: Qwen2-VL-7B-Instruct reaches $\sim$28\%, while LLaVA fails to generalize (>45\%). A diagram rarely helps: only Gemini shows clear gains, suggesting its stronger capacity for abstract visual reasoning. 

Example-based prompts substantially improve Qwen's performance. A simple CLIP-based kNN baseline without VLMs yields $\sim$21\% error, while conditioning Qwen2-VL-7B-Instruct on the same retrieved samples can lower error to $\sim$16\% (Fig.~\ref{fig:knn_comparison}) and Gemini to $\sim$9\%. Balanced retrieval for Qwen degrades this result (to $\sim$27\%), Gemini scores at best $\sim$11\%, highlighting its potential for visual reasoning from examples. However, it also shows greater variance across prompt rewordings. This indicates that strong performance may arise from incidental prompt alignment rather than reasoning.

CoT generally increases variance across prompts and, on average, performs worse across models. However, in some cases, it uncovers strong reasoning paths that lead to significantly better performance, such as in Gemini under Text and kNN-Balanced settings. These results suggest that while CoT has potential, its effective use currently requires substantial supervision and careful prompt testing.

GPT-4o, tested on a subset of prompts, underperforms Gemini and at times Qwen2, highlighting that even top proprietary VLMs can struggle with domain-specific visual tasks. In our experiments GPT-4o-mini performs better than GPT-4o and achieves $\sim$36--38\% error in Text and Diagram settings, and $\sim$22--34\% across example-based prompts.

\paragraph{Sensitivity.}
We investigate prompt sensitivity using Qwen2-VL-7B-Instruct, as it achieved the strongest performance among open-source models. As shown Fig.~\ref{fig:temperature_influence}, lower temperatures yield more stable results. Changing the order of support images (Fig.~\ref{fig:permutations_instability}) shifts error by up to 10 points likely due to positional attention biases, where earlier tokens/images dominate the model’s focus. This underlines a critical limitation: even strong VLMs may rely on shallow heuristics, making their behavior fragile and hard to trust without careful prompt control.

\paragraph{Supervised fine-tuning.}
Finally, we fine-tune Qwen2-VL-7B-Instruct using LoRA, updating $\sim$15M parameters (see Table~\ref{tab:mirabest_confident}). With 729 samples, test error drops to 3.1\%, close to AstroVFM’s 1.9\%~\cite{slijepcevic2024radio}, despite lacking astronomy-specific pretraining. Using 145 labels, Qwen2 starts outperforming a ResNet trained from scratch~\cite{slijepcevic2024radio} and consistently strengthens its lead as data grow. This shows that VLMs can rival domain-specialized models with lightweight adaptation, offering a scalable and data-efficient alternative, though task-specific architectures are still the best choice.

\begin{table}[t]
  \centering
  \small
\caption{Test-set error on MiraBest with increasing sample counts. We compare Qwen2-VL-7B-Instruct fine-tuned with LoRA against training from scratch and AstroVFM results reported in~\cite{slijepcevic2024radio}. Bold values indicate the best performance. All results are averaged over 10 runs per setting. }
\begin{tabular}{rccc}
    \toprule
    Samples & ResNet~\cite{slijepcevic2024radio} & AstroVFM~\cite{slijepcevic2024radio} & Qwen2-VL + LoRA \\
    \midrule
     36  & 19.0 $\pm$ 1.4  & \textbf{16.2 $\pm$ 1.0} & 24.0 $\pm$ 2.5 \\
     72  & 18.8 $\pm$ 0.7  & \textbf{11.8 $\pm$ 0.1} & 16.2 $\pm$ 3.5 \\
    145  & 11.9 $\pm$ 0.5  & \textbf{7.8  $\pm$ 0.8} & 10.8 $\pm$ 2.5 \\
    291  &  7.0 $\pm$ 0.6  & \textbf{4.9  $\pm$ 0.5} &  6.9 $\pm$ 1.7 \\
    437  &  5.8 $\pm$ 0.5  & \textbf{4.5  $\pm$ 0.5} &  5.6 $\pm$ 1.6 \\
    583  &  5.1 $\pm$ 0.3  & \textbf{4.3  $\pm$ 0.4} &  4.6 $\pm$ 1.5 \\
    656  &  4.8 $\pm$ 0.2  & \textbf{3.1  $\pm$ 0.4} &  3.3 $\pm$ 1.4 \\
    729  &  4.8 $\pm$ 0.2  & \textbf{1.9  $\pm$ 0.3} &  3.1 $\pm$ 1.0 \\
    \bottomrule
  \end{tabular}
  \label{tab:mirabest_confident}
\end{table}

\section{Discussion and Conclusion}
\label{sec:colclusions}
We investigate the capabilities and limitations of VLMs on the MiraBest radio galaxy dataset, comparing five prompting strategies across open and proprietary models. Our results reveal that zero-shot classification is feasible: Gemini achieves strong performance ($\sim$14\% error) using only text prompts, while Qwen2 models can approach it when conditioned on visual examples. Example-based conditioning is very effective for open models. We show that Qwen2-VL-7B-Instruct outperforms a CLIP-based $k$NN classifier by $\sim$5 points when prompted with retrieved images and labels. We also introduce a balanced retrieval setup, less explored in prior work, which leads to distinct behaviors. Notably, Gemini with noCoT performs well under this regime, with slightly increased variance, while others degrade significantly, suggesting stronger abstraction for Gemini. CoT prompts show mixed effects: while they perform well on certain prompts, they also introduce instability and increase variance, highlighting limitations in current VLM reasoning on scientific tasks.

Fine-tuning with LoRA allows generic VLMs to rival astronomy-specific models. With 145 labeled examples, Qwen2 outperforms a scratch-trained ResNet on the same subset; with 729 labels, it reaches 3.1\% test error, which is near AstroVFM’s 1.9\% despite no domain-specific pretraining.

Overall, our findings encourage the use of VLMs in scientific imaging, while underscoring the need for carefully designed prompts and adaptation strategies.

\section{Acknowledgements}

M. Drozdova and V. Kinakh are supported by the RODEM: Robust deep density models for high-energy particle physics and solar flare analysis Sinergia Project funded by the Swiss National Science Foundation, grant number CRSII5-193716.

\section{Reproducibility}
\label{app:reproducability}
All source code and resources for our experiments are publicly available at \url{https://github.com/MariiaDrozdova/application_VLM_to_astronomy}.

\bibliographystyle{plain} 
\bibliography{main}

\newpage
\appendix

\section{Dataset and models}
\label{app:dataset}

\begin{figure}[t]
    \centering
    \begin{subfigure}[b]{0.3\linewidth}
        \centering
        \includegraphics[width=\linewidth]{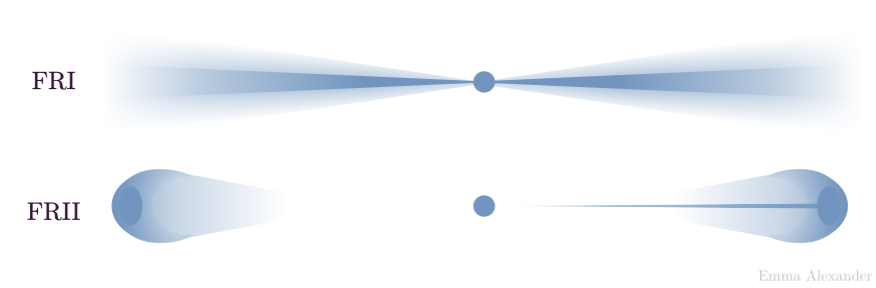}
        \caption{Diagram}
        \label{fig:fr_schematic}
    \end{subfigure}
    \hfill
    \begin{subfigure}[b]{0.28\linewidth}
        \centering
        \includegraphics[width=\linewidth]{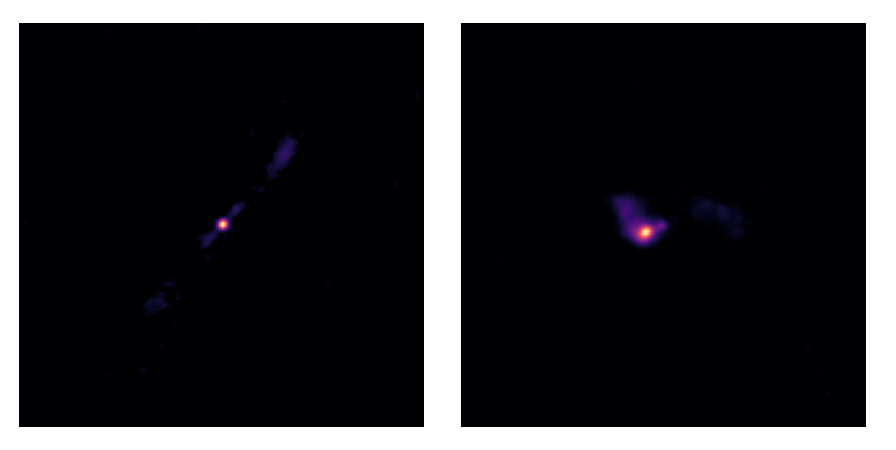}
        \caption{FR-I}
        \label{fig:fr_examples1}
    \end{subfigure}
    \hfill
    \begin{subfigure}[b]{0.28\linewidth}
        \centering
        \includegraphics[width=\linewidth]{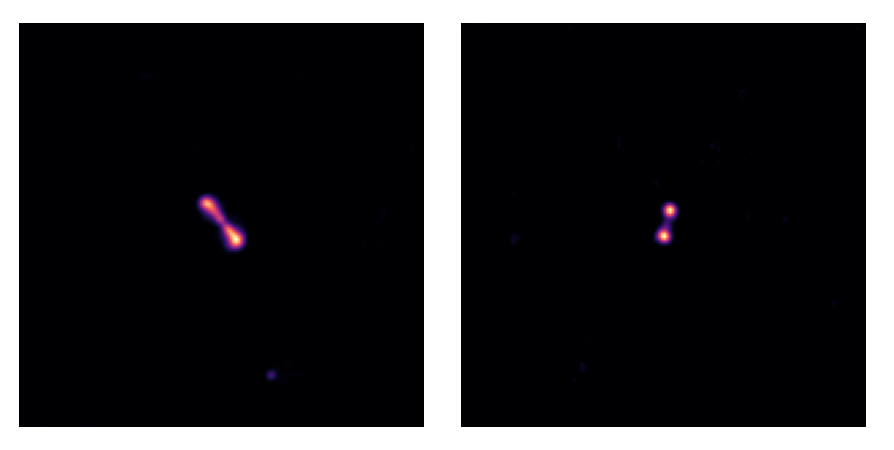}
        \caption{FR-II}
        \label{fig:fr_examples2}
    \end{subfigure}
    \caption{
    FR-I vs FR-II radio galaxy morphologies. (a) Schematic illustration of the Fanaroff–Riley classification \cite{alexander_resources}.  (b–c) MiraBest radio images.}
    \label{fig:fr_combined}
\end{figure}

\paragraph{Dataset.}
We study morphology-based binary classification of radio galaxies into Fanaroff–Riley types I and II~\cite{fanaroff1974morphology}. FR-I galaxies have bright central cores with fading jets (Fig.~\ref{fig:fr_examples1}); FR-II exhibit edge-brightened lobes with prominent hotspots (Fig.~\ref{fig:fr_examples2}). We use the \texttt{MiraBest-confident} split~\cite{porter2023mirabest}, a class-balanced set of 729 training and 104 test images from NVSS~\cite{condon1998nrao} and FIRST~\cite{becker1994vla}, labeled by experts. Although public and potentially seen during VLM pretraining, we treat this task effectively out-of-domain relative to typical natural image distributions.

\paragraph{Models.}
We evaluate open-source VLMs, such as Qwen2/2.5-VL-Instruct~\cite{team2024qwen2}, 2B–7B; LLaVA-1.6-Mistral~\cite{liu2024llavanext}, using chat-style prompting with structured system/user/assistant roles. Proprietary models, such as Gemini-2.5-Flash~\cite{comanici2025gemini} and GPT-4o~\cite{openai_chatgpt_2023}, are accessed via API with fixed decoding settings; both use conversational interfaces.

\include{project/average_runtime}
\include{project/detailed_results}
\include{project/finetuning}
\include{project/text_prompts}
\include{project/diagram_prompts}
\include{project/few_shot_prompt}
\include{project/extra_tables}
\end{document}

%% file: project/average_runtime.tex
\section{Detailed Runtime Analysis}
\label{app:runtime}

We report detailed inference runtimes across input types and prompting strategies for each model (Table~\ref{tab:runtime_cot_makecell}). Each cell shows the mean and standard deviation of per-sample response time (in milliseconds) under two prompting regimes: without CoT and with CoT.

As expected, CoT prompting increases inference time quite substantially. LLaVA and Qwen-based models for example, have 5-10× difference between noCoT/CoT. Gemini models, in contrast, exhibit more stable behavior, with only modest increases in runtime. However, Gemini's average latency remains high due to API throttling or availability delays, which we observed during evaluation.

Input modality also impacts runtime: prompts with retrieved visual examples (e.g., kNN-Imgs) takes on average more time.

\begin{table*}[ht]
\centering
\small
\begin{tabular}{lccccc}
\toprule
\textbf{Model} & \textbf{Text} & \textbf{Diagram} & \textbf{Fixed-Imgs} & \textbf{kNN-Imgs} & \textbf{kNN-Balanced} \\
\midrule
Gemini & \makecell{526 ± 341\\549 ± 231} & \makecell{880 ± 25\\1100 ± 95} & \makecell{898 ± 11\\1035 ± 95} & \makecell{889 ± 16\\1022 ± 83} & \makecell{860 ± 1\\1030 ± 82} \\
\hdashline
LLaVA-1.6 & \makecell{72 ± 28\\317 ± 179} & \makecell{117 ± 23\\497 ± 231} & \makecell{68 ± 6\\510 ± 140} & \makecell{133 ± 44\\1246 ± 1663} & \makecell{179 ± 104\\608 ± 205} \\
\hdashline
Qwen2-2B & \makecell{8 ± 3\\79 ± 36} & \makecell{10 ± 4\\58 ± 42} & \makecell{10 ± 4\\34 ± 47} & \makecell{24 ± 6\\45 ± 37} & \makecell{22 ± 4\\43 ± 39} \\
\hdashline
Qwen2-7B & \makecell{11 ± 14\\246 ± 397} & \makecell{11 ± 2\\121 ± 57} & \makecell{11 ± 2\\102 ± 99} & \makecell{33 ± 8\\152 ± 153} & \makecell{30 ± 6\\142 ± 126} \\
\hdashline
Qwen2.5-3B & \makecell{6 ± 2\\130 ± 30} & \makecell{10 ± 2\\70 ± 34} & \makecell{8 ± 2\\192 ± 143} & \makecell{22 ± 1\\172 ± 120} & \makecell{20 ± 1\\173 ± 114} \\
\hdashline
Qwen2.5-7B & \makecell{8 ± 3\\169 ± 81} & \makecell{13 ± 2\\241 ± 83} & \makecell{13 ± 3\\326 ± 272} & \makecell{24 ± 0\\263 ± 172} & \makecell{22 ± 0\\262 ± 180} \\
\bottomrule
\end{tabular}
\caption{
\textbf{Average per-sample inference time (ms)} across models and prompt formats. 
Each cell reports results for \textit{noCoT} (top) and \textit{with CoT} (bottom). 
LLaVA and Qwen variants show significant runtime increases with CoT prompts. Gemini runtimes are more stable, though influenced by occasional API throttling.
}
\label{tab:runtime_cot_makecell}
\end{table*}

%% file: project/detailed_results.tex
\newlength{\panelht}
\setlength{\panelht}{5.9cm} 
\newlength{\panelhtzs}
\setlength{\panelhtzs}{\panelht} 
\newlength{\panelhtfs}
\setlength{\panelhtfs}{\panelht} 
\newlength{\panelhtfsknn}
\setlength{\panelhtfsknn}{\panelht} 
\newlength{\panelhtfsknnb}
\setlength{\panelhtfsknnb}{\panelht} 

\section{Full Prompting Results}
\label{app:analysis}

This appendix contains detailed results for each prompting strategy evaluated in our study. All evaluations use greedy decoding. We test placing the query image before and after the text query. A mapping of CoT versus NoCoT prompts from Appendix \ref{app:prompts} is provided in Table~\ref{tab:cot-indexes}.

\begin{table}[h]
\centering
\caption{Chain-of-Thought (CoT) and NoCoT Prompt Indexes by Strategy}
\label{tab:cot-indexes}
\begin{tabular}{l|p{5cm}|p{5cm}}
\toprule
\textbf{Strategy} & \textbf{CoT Prompt Indexes} & \textbf{NoCoT Prompt Indexes} \\
\midrule
Text      & 3, 7, 8, 9, 10, 12 & 0, 1, 2, 5, 6, 11 \\
Diagram   & 2, 3, 4, 5, 6      & 0, 1, 7 \\
Few-shot  & 2, 3, 4, 5         & 0, 1, 6, 7 \\
\bottomrule
\end{tabular}
\end{table}

\paragraph{Prompt Formats.}
We employ five distinct prompting strategies: \texttt{Text}, \texttt{Diagram}, \texttt{Fixed-Imgs}, \texttt{kNN-Imgs} and \texttt{kNN-Balanced}. Each strategy is constructed using structured system/user/assistant roles, with the following component orderings (components are clarified in Appendix \ref{app:prompts}):

\begin{itemize}
    \item \textbf{Text.} The \texttt{system\_message} is always provided first. This is followed by either:
    \begin{enumerate}
        \item the \texttt{query\_text} and then the \texttt{query\_image}, or
        \item the \texttt{query\_image} and then the \texttt{query\_text},
    \end{enumerate}
    depending on the layout setting being evaluated.

    \item \textbf{Diagram.} The schematic diagram is always placed \textit{first} in the prompt. It is followed by the \texttt{system\_message}, and then the \texttt{query\_text} and \texttt{query\_image}, in the order depending on the layout setting being evaluated.

    \item \textbf{Fixed-Imgs/kNN-Imgs/kNN-Balanced} The prompt begins with the \texttt{system\_message}, followed by an optional \texttt{examples\_message} that introduces the few-shot exemplars (this is typically left empty). This is followed by a set of few-shot (user, assistant) pairs, where:
    \begin{itemize}
        \item The user message contains a tuple: (\texttt{query\_text\_example}, \texttt{image\_example}).
        \item The assistant message provides the corresponding \texttt{label}.
    \end{itemize}
    The order of elements in each (\texttt{query\_text\_example}, \texttt{image\_example}) tuple matches that of the final query (\texttt{query\_text}, \texttt{query\_image}) and depends on the specified layout (image-first or image-last).
\end{itemize}

In several prompts, we enforce a structured response format using \texttt{<think>}...\texttt{</think>} and \texttt{<answer>}...\texttt{</answer>} tags. However, open-source models frequently fail to consistently adhere to this format. As a result, rather than relying on regular expression-based parsing, we ultimately adopted a simpler heuristic: extracting the last class label mentioned in the output.

\begin{figure}[t]
  \centering
  \begin{subfigure}[t]{0.31\linewidth}
    \centering
    \includegraphics[height=\panelhtzs,trim=0 0 80bp 0,clip]{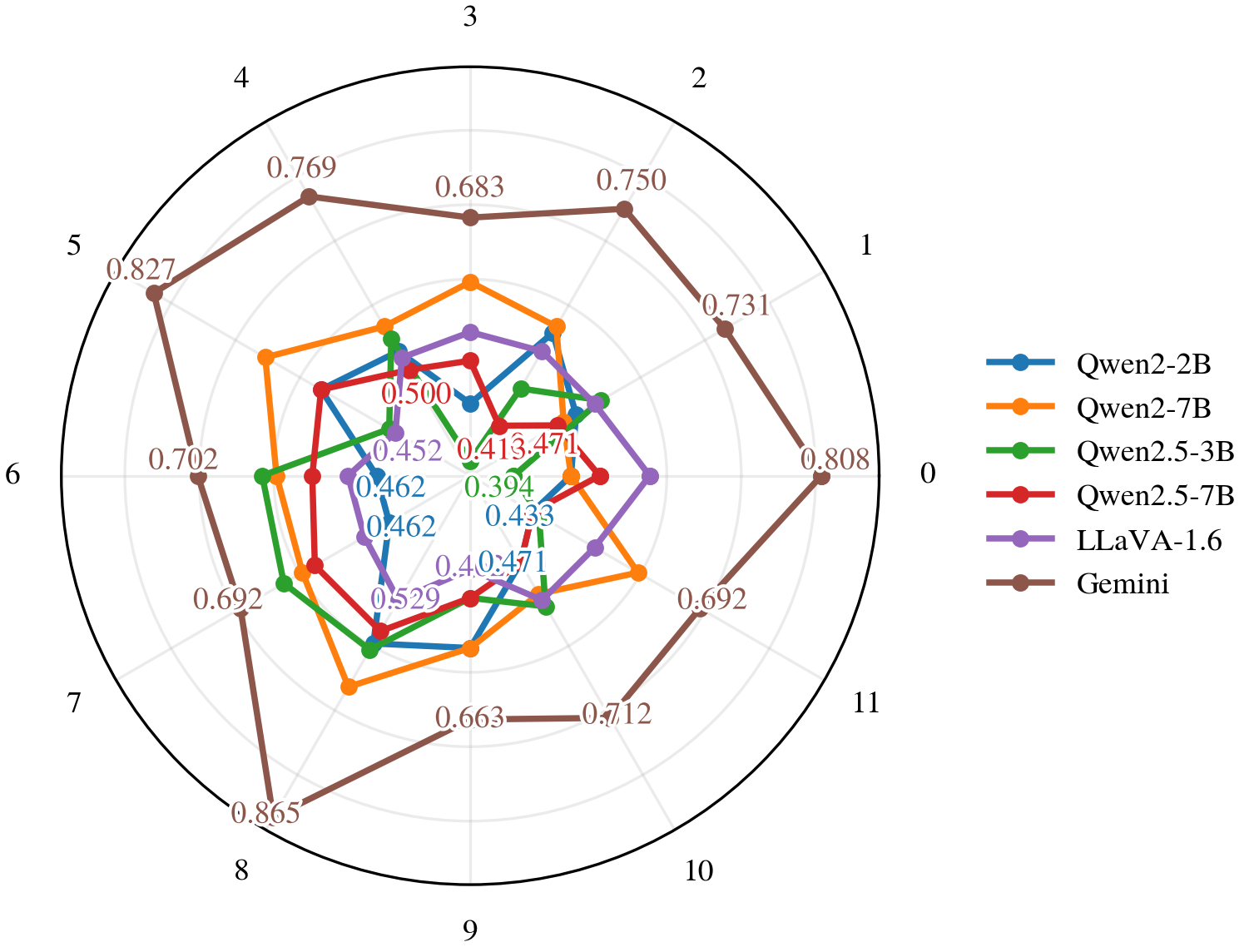}
    \subcaption{}
  \end{subfigure}\hfill
  \begin{subfigure}[t]{0.61\linewidth}
    \centering
    \includegraphics[height=\panelhtzs]{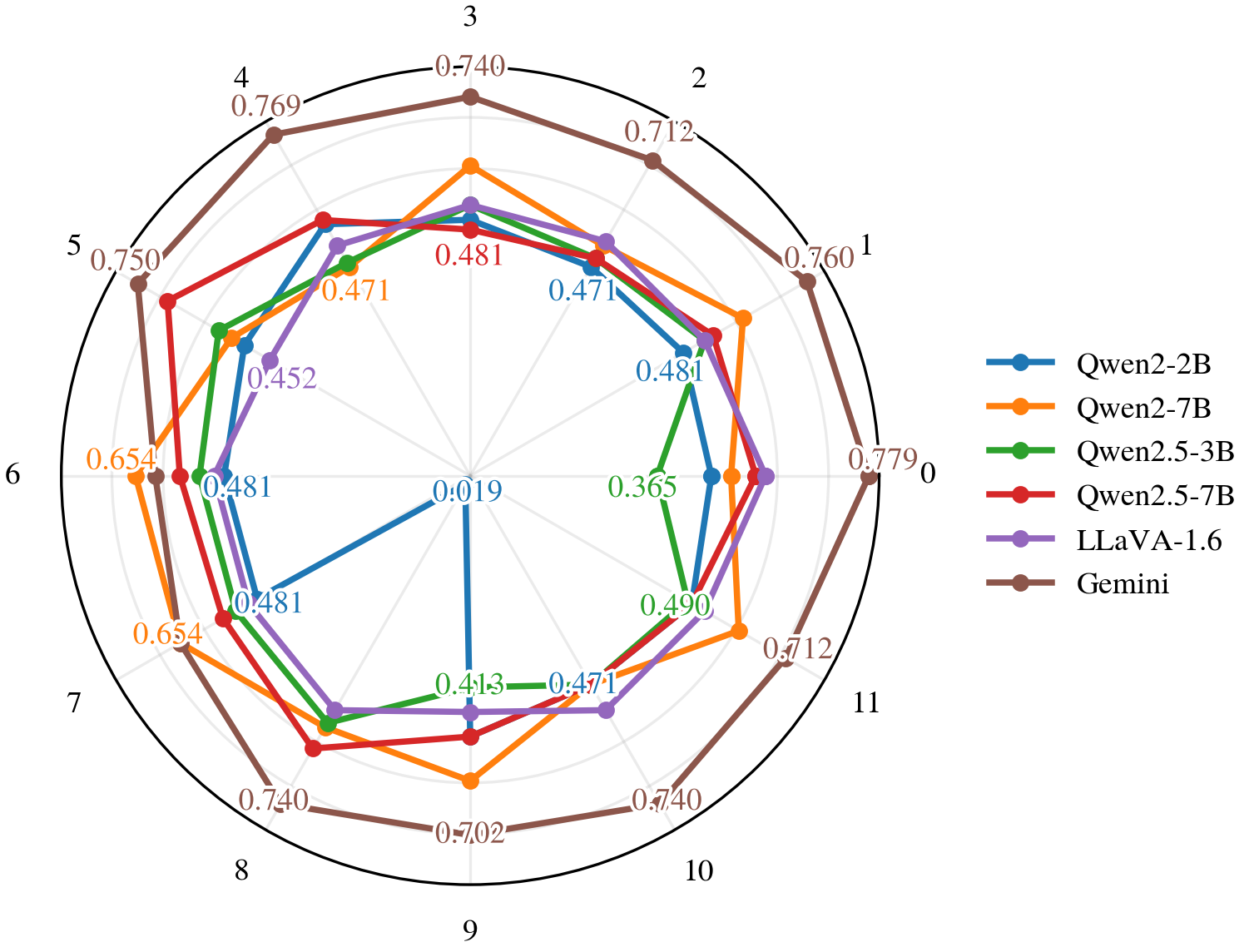}
    \subcaption{}
  \end{subfigure}
  \caption{Radar plots showing test accuracy across Text prompts. (a) Image placed after the query; (b) image placed before. Gemini consistently outperforms other models across prompt variants.}
  \label{fig:radar1}
\end{figure}

\begin{figure}[t]
  \centering
  \begin{subfigure}[t]{0.31\linewidth}
    \centering
    \includegraphics[height=\panelhtzs,trim=0 0 80bp 0,clip]{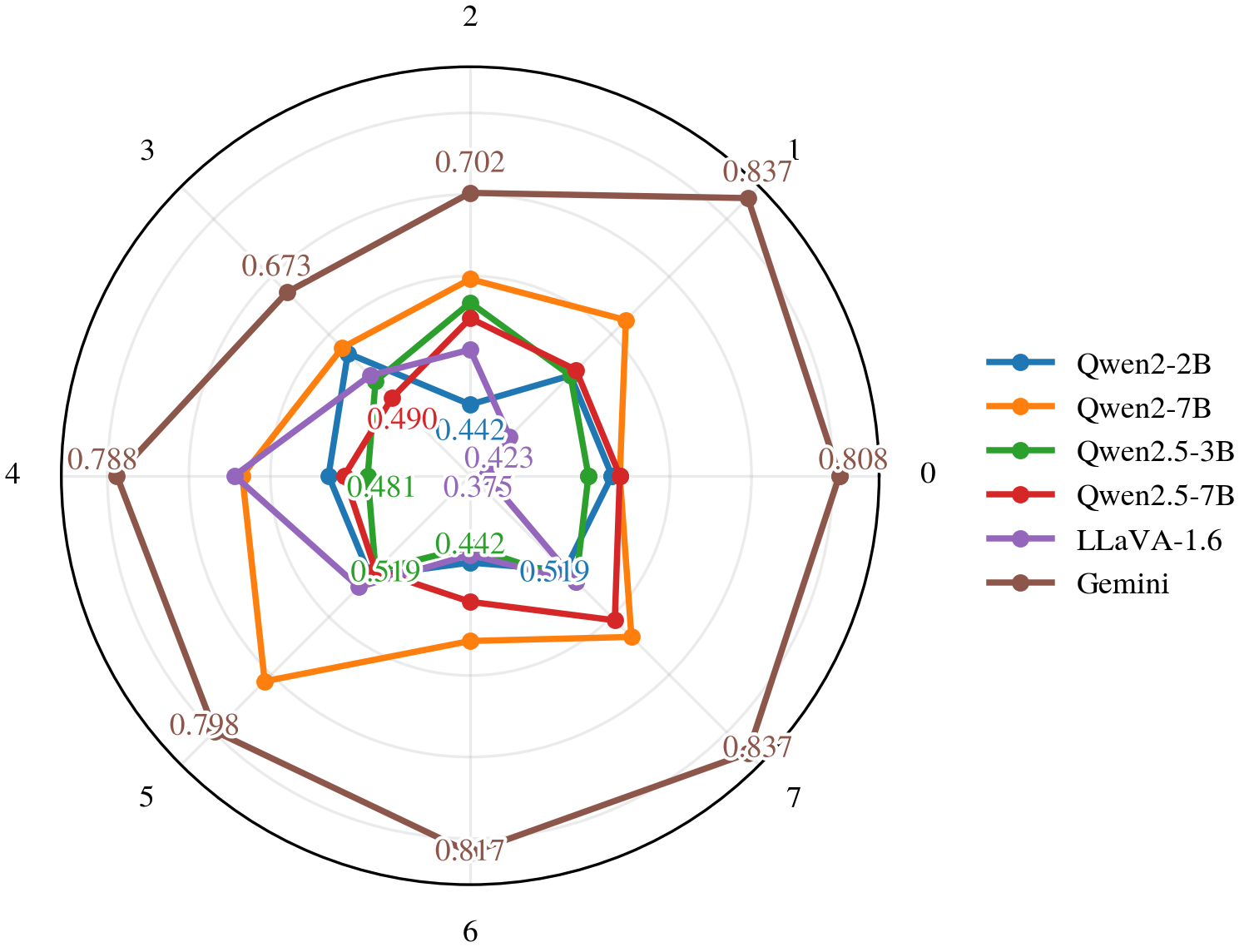}
    \subcaption{}
  \end{subfigure}\hfill
  \begin{subfigure}[t]{0.61\linewidth}
    \centering
    \includegraphics[height=\panelhtzs]{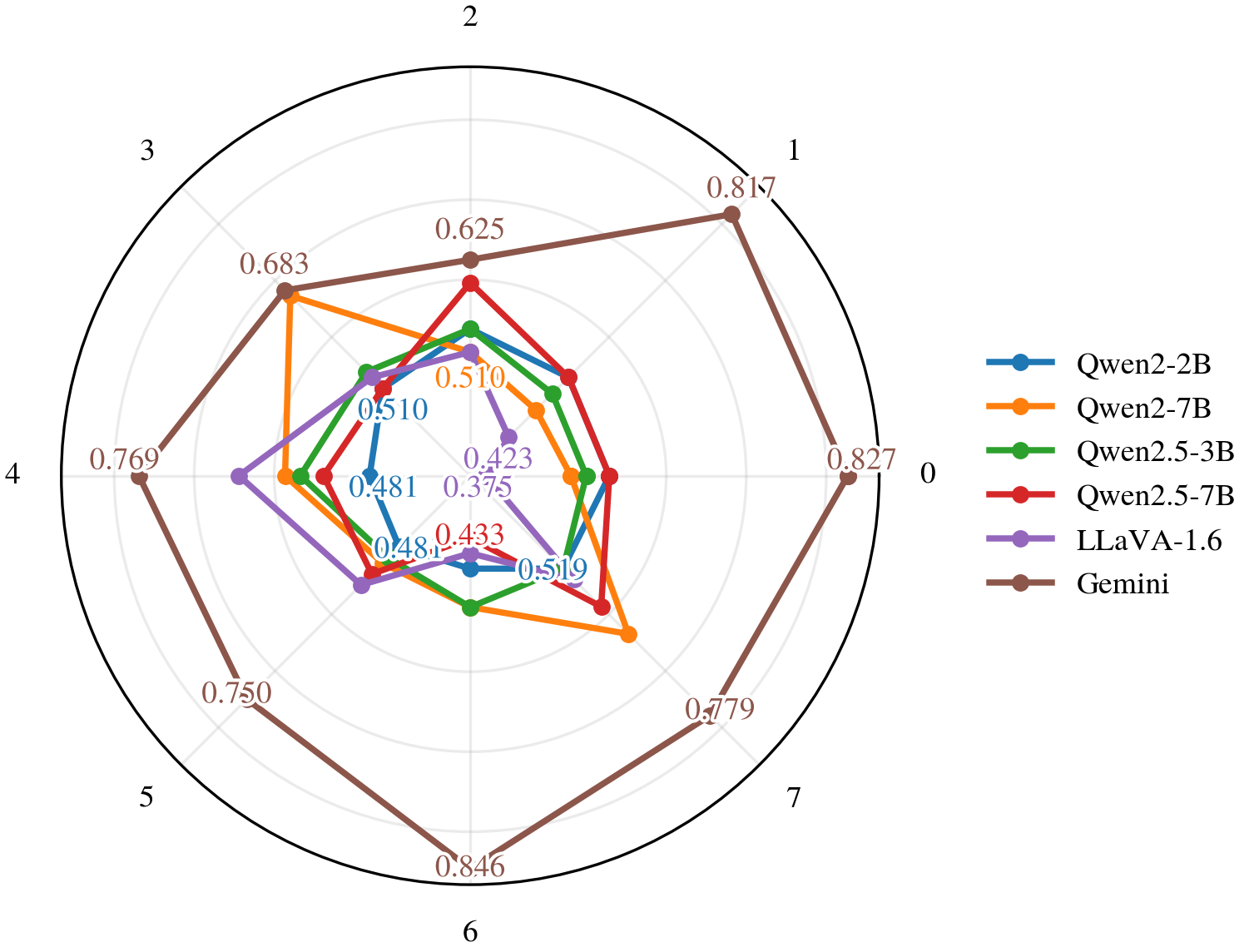}
    \subcaption{}
  \end{subfigure}
  \caption{Radar plots showing test accuracy across Diagram prompts. (a) Image placed after the query; (b) image placed before. Gemini consistently outperforms other models across prompt variants.}
  \label{fig:radar2}
\end{figure}

\begin{figure}[t]
  \centering
  \begin{subfigure}[t]{0.31\linewidth}
    \centering
    \includegraphics[height=\panelhtzs,trim=0 0 80bp 0,clip]{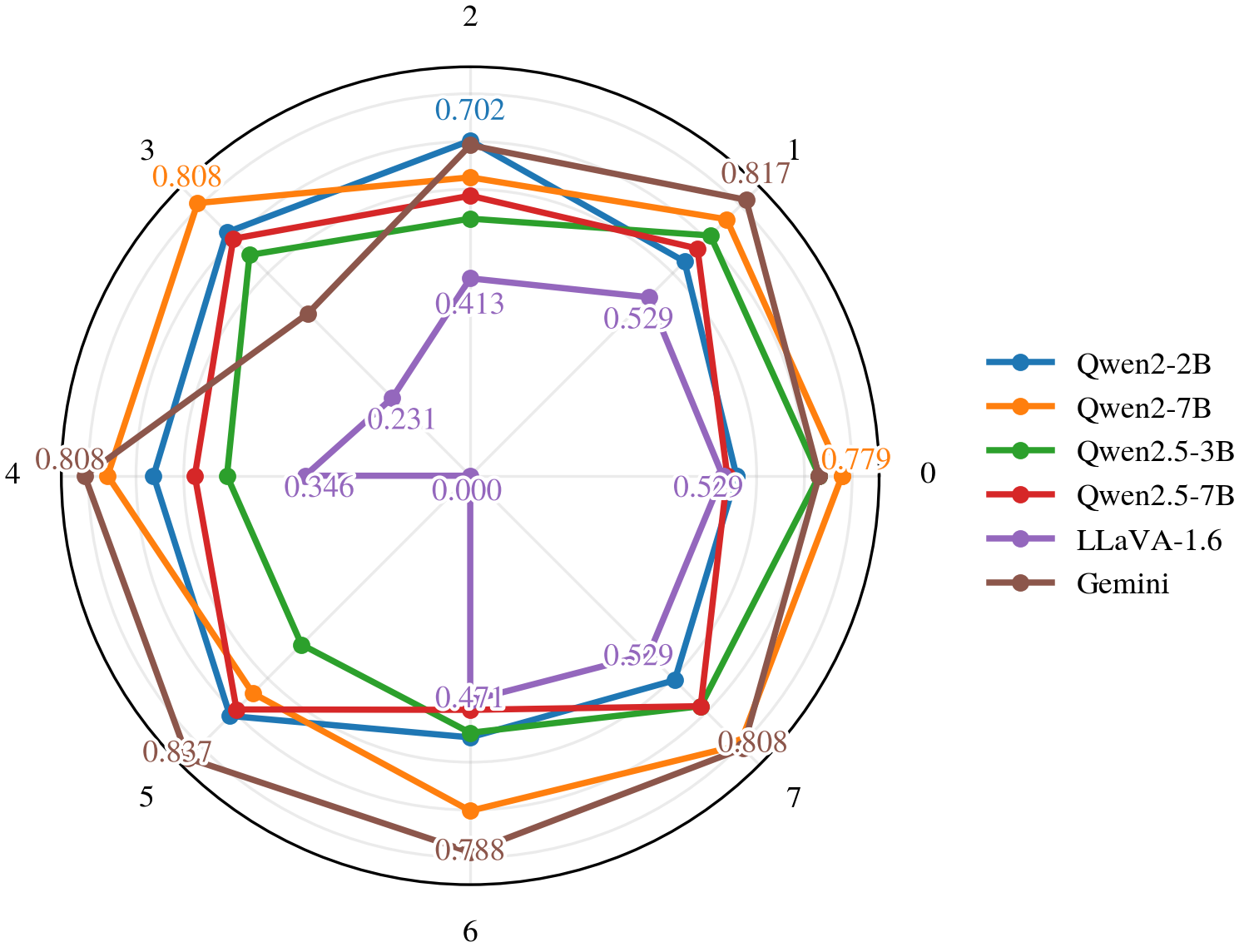}
    \subcaption{}
  \end{subfigure}\hfill
  \begin{subfigure}[t]{0.61\linewidth}
    \centering
    \includegraphics[height=\panelhtzs]{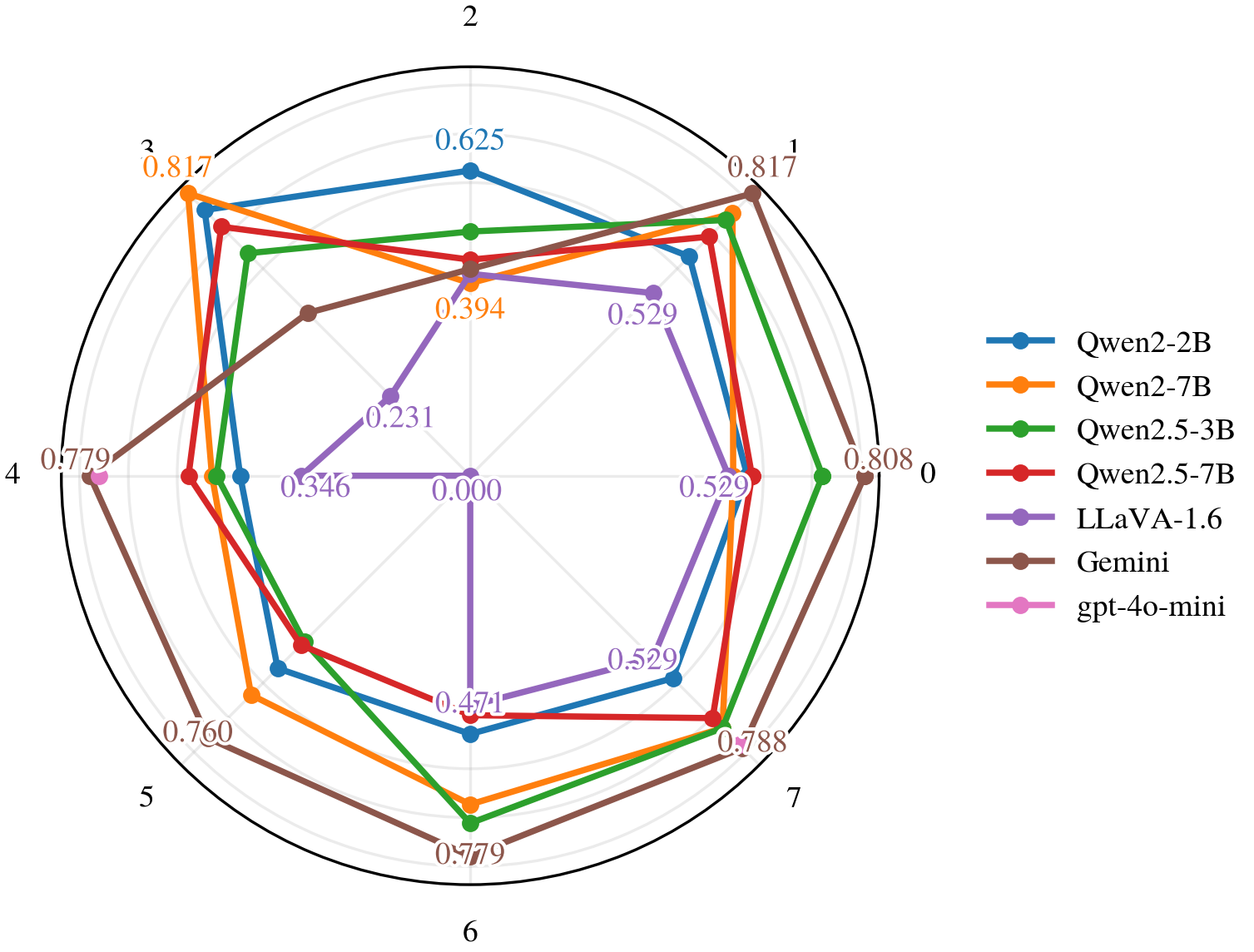}
    \subcaption{}
  \end{subfigure}
  \caption{Radar plots showing test accuracy across Fixed-Imgs prompts. (a) Image placed after the query; (b) image placed before. Open models performance is considerably improved. }
  \label{fig:radar3}
\end{figure}

\begin{figure}[t]
  \centering
  \begin{subfigure}[t]{0.31\linewidth}
    \centering
    \includegraphics[height=\panelhtzs,trim=0 0 80bp 0,clip]{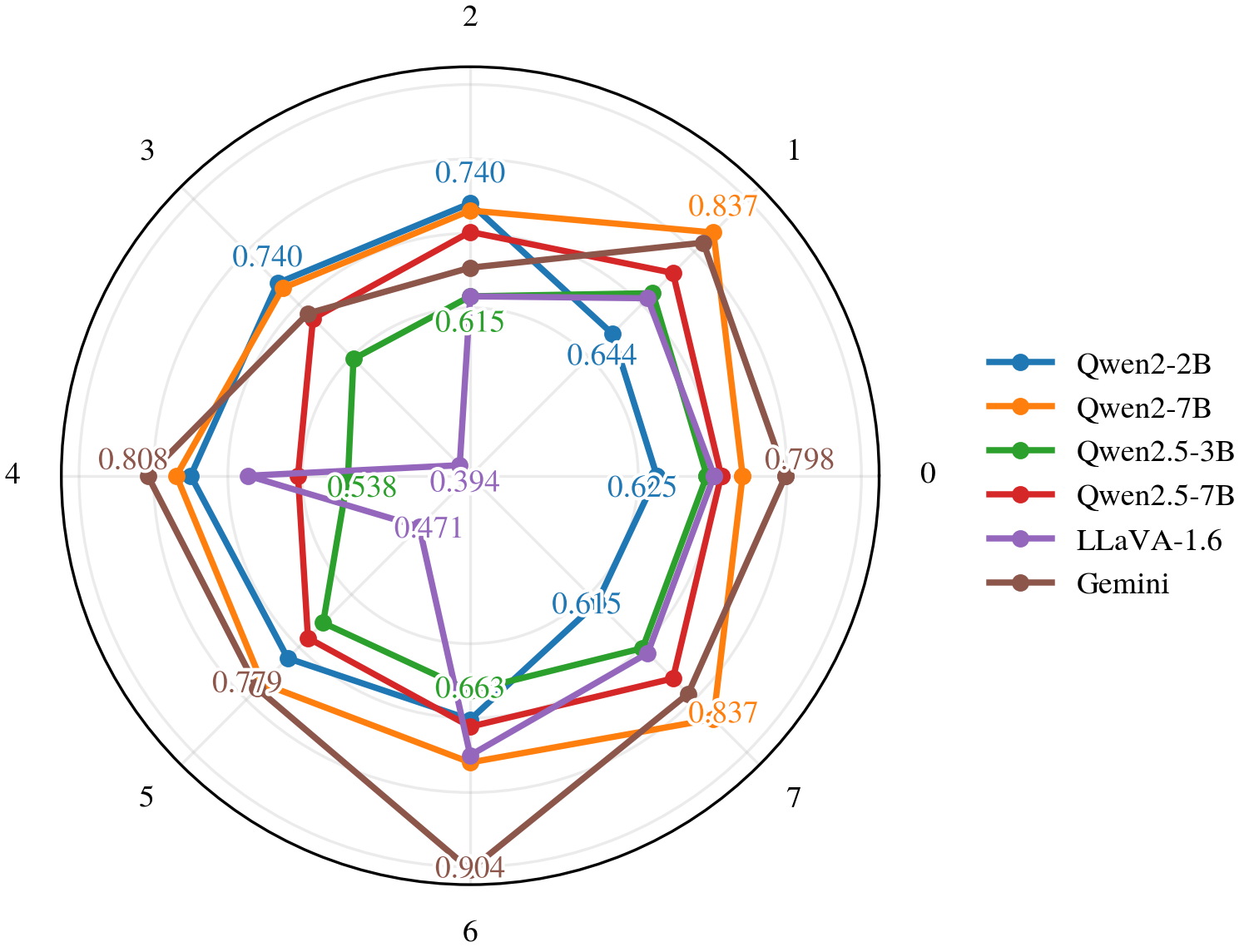}
    \subcaption{}
  \end{subfigure}\hfill
  \begin{subfigure}[t]{0.61\linewidth}
    \centering
    \includegraphics[height=\panelhtzs]{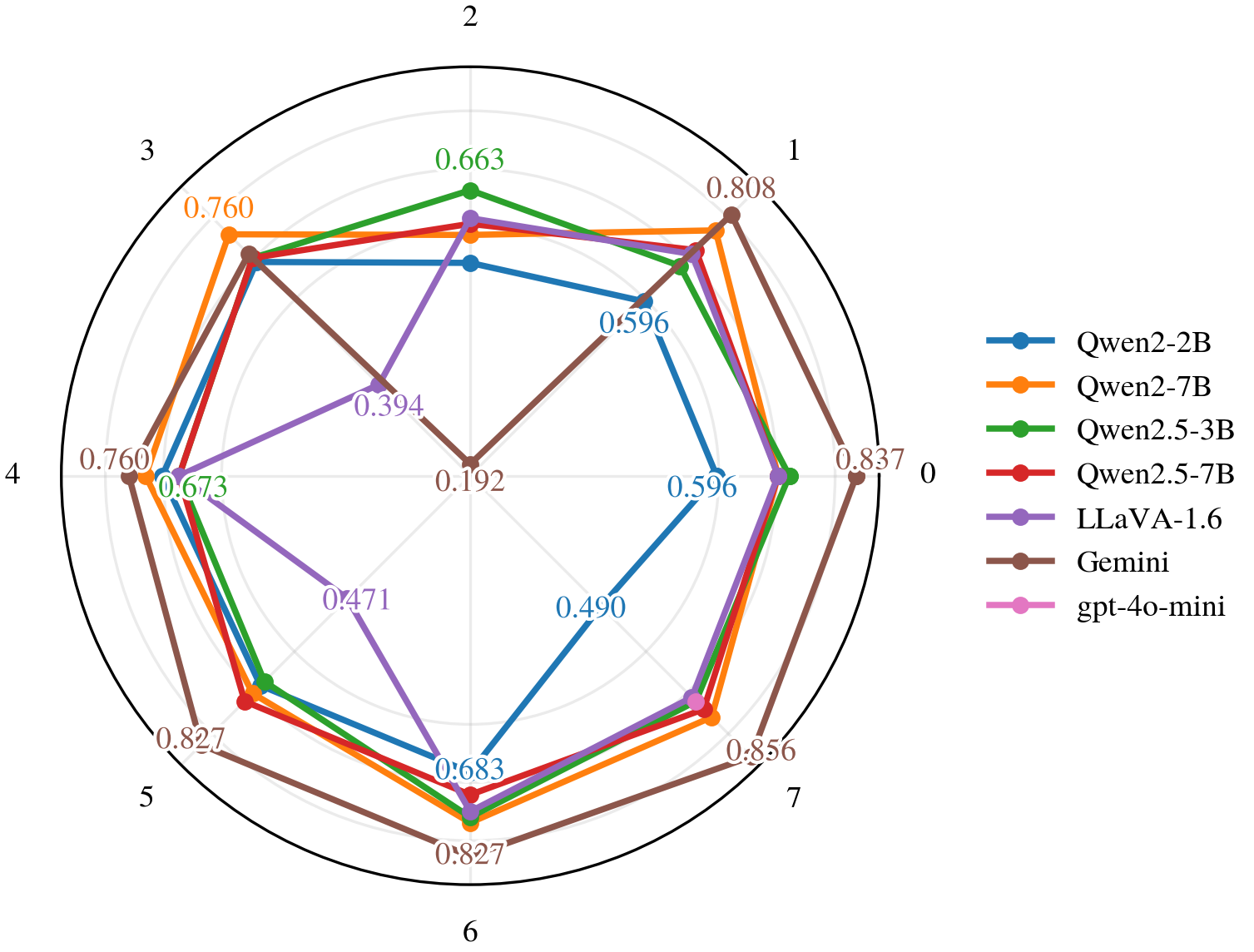}
    \subcaption{}
  \end{subfigure}
  \caption{Radar plots showing test accuracy across kNN-Imgs prompts. (a) Image placed after the query; (b) image placed before. Open models perform well, especially Qwen2-VL-7B-Instruct. Gemini achieves its best result (prompt 6).}
  \label{fig:radar4}
\end{figure}

\begin{figure}[t]
  \centering
  \begin{subfigure}[t]{0.31\linewidth}
    \centering
    \includegraphics[height=\panelhtzs,trim=0 0 80bp 0,clip]{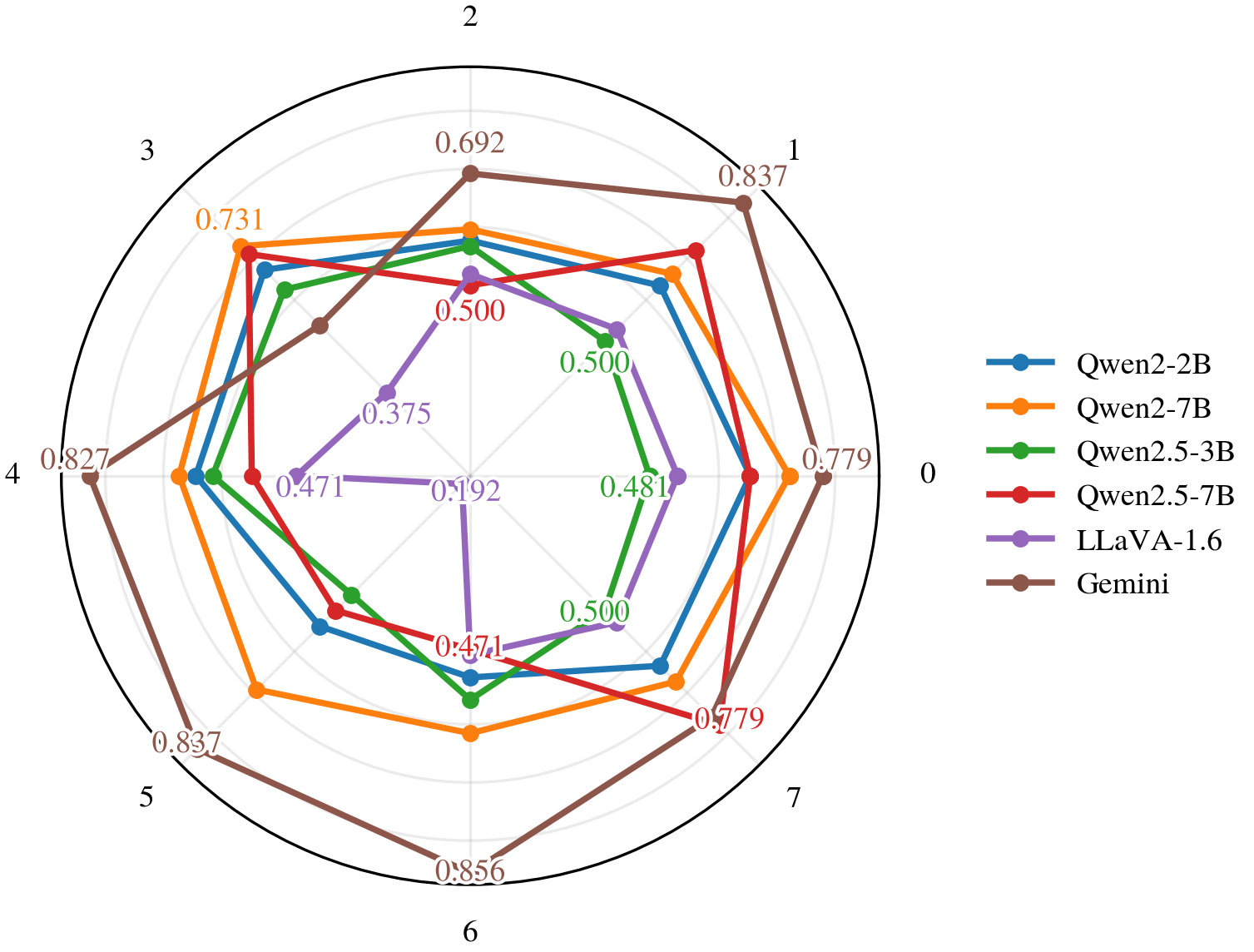}
    \subcaption{}
  \end{subfigure}\hfill
  \begin{subfigure}[t]{0.61\linewidth}
    \centering
    \includegraphics[height=\panelhtzs]{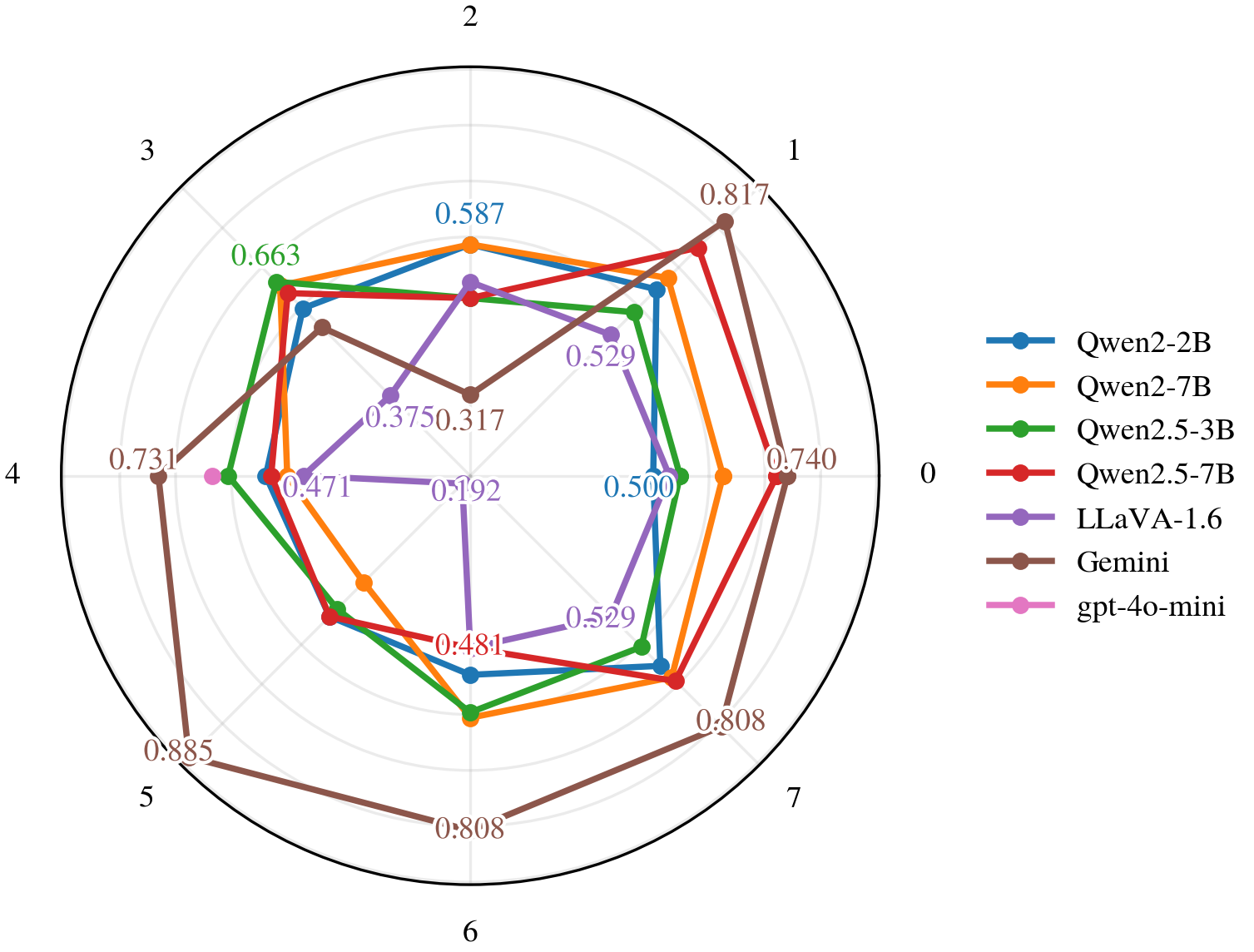}
    \subcaption{}
  \end{subfigure}
  \caption{Radar plots showing test accuracy across kNN-Balanced prompts. (a) Image placed after the query; (b) image placed before. Performance drops for open models compared to kNN-Imgs. Gemini remains more robust.}
  \label{fig:radar5}
\end{figure}

\paragraph{Visual Summaries.}
The radar plots provided in Figures~\ref{fig:radar1}\ref{fig:radar2}\ref{fig:radar3}, \ref{fig:radar4}, \ref{fig:radar5} offer a visual summary of prompt-wise performance (accuracy). For reference:
\begin{itemize}
    \item Text prompt results are visualized in Figure~\ref{fig:radar1} and in Tables \ref{tab:text_0}, \ref{tab:text_1}.
    \item Diagram prompt results are shown in Figure~\ref{fig:radar2} and in Tables \ref{tab:diagram_0}, \ref{tab:diagram_1}.
    \item Few-shot prompts (Fixed-Imgs, kNN-Imgs, kNN-Balanced) are summarized in Figures~\ref{fig:radar3}, \ref{fig:radar4}, and \ref{fig:radar5}, respectively.
    For Fixed-Imgs Tables \ref{tab:fixed_imgs_0}, \ref{tab:fixed_imgs_1}, for kNN-Imgs Tables\ref{tab:knn_imgs_0}, \ref{tab:knn_imgs_1} and for kNN-Balanced Tables \ref{tab:knn_balanced_0}, \ref{tab:knn_balanced_1}.
\end{itemize}

In the tables \texttt{index} corresponds a prompt number from Appendix \ref{app:prompts}.

\subsection*{Prompt-Level Analysis and Observations}

In this section, we highlight a few cases that offer insights into the impact of prompt formulation.

\paragraph{Gemini}
The most striking variation was observed in Gemini's performance under the kNN-Imgs setup. In particular, for prompt 2, Gemini got 81\% error, while prompt 6 yielded just 9\%, a dramatic 72-point gap under otherwise comparable conditions. Prompt 2 implies chain-of-thought reasoning, instructing the model to respond using a \texttt{<think>}...\texttt{</think>} block followed by an \texttt{<answer>}...\texttt{</answer>} block and one more request to explain motivation and reasoning. This request in the end leads Gemini to often continue generating after the \texttt{<answer>}...\texttt{</answer>} block, finishing with phrases of type “it is definitely not FR-II.” Our heuristic extracts the last mentioned class (e.g., “FR-II” in the above case), which leads to incorrect labeling.  This explains bad results for prompt 2 in Gemini's case.

\paragraph{General Trends Across Open Models.}
For open-source models, especially Qwen variants, prompts that avoid asking for extended reasoning tend to perform better. In other words, shorter, direct prompts yield more reliable and accurate results. Prompt 2 often underperforms across models in few-shot settings, likely for the same reasons discussed above.

\paragraph{Probabilistic Reasoning Helps in Gemini.}
Interestingly, prompt 8 performed particularly well for Gemini in the zero-shot text-only setting. This prompt first instructs the model to list relevant visual features, then analyze them, and finally estimate the probability (from 0 to 1) of each class before concluding with a final decision. This structured probabilistic reasoning appears to improve performance for Gemini. However, the same prompt does not show consistent benefits for most open models, although Qwen2-VL-7B-Instruct also achieves its best score among text-only prompts (33.7\%).

\paragraph{Model-Specific Observations.}
Across all prompting strategies, Qwen2-VL-7B-Instruct outperforms not only smaller Qwen models and LLaVa but even Qwen2.5-7B-Instruct(except no CoT in kNN-Balanced). While both are similar in scale, Qwen2 variant appears better aligned with general reasoning tasks. One possible explanation is that Qwen2.5-7B, being more specialized or fine-tuned for specific downstream tasks, may have narrower behavior, whereas Qwen2-VL-7B-Instruct retains broader generalization capabilities. However, this remains speculative and requires further investigation.

%% file: project/finetuning.tex
\section{Notes on LoRA Fine-Tuning}
\label{app:lora}

We fine-tune Qwen2-VL-7B-Instruct using LoRA~\cite{hu2022lora}, updating only $\sim$15M parameters across both the vision and language modules.

Instead of fine-tuning on image–label pairs alone, we include the full prompt during training. As in zero-shot inference, we prepend a system message describing the FR-I and FR-II classes, followed by a direct classification question. The exact prompt used was:

\begin{nolinenumbers}
\begin{promptlisting}
system_message = (
            "You are an expert radio galaxy classifier. "
            "FR-I: bright core, gradually fading jets. "
            "FR-II: faint core, jets end in hotspots. "
            "Consider core brightness and jet termination."
            )
query_text = 'Respond only FR-I or FR-II.'
\end{promptlisting}
\end{nolinenumbers}

This corresponds to Prompt 0 in our zero-shot setup (see Appendix~\ref{app:prompts}). We did not experiment with other zero-shot prompt variants for training.

We observed that LoRA fine-tuning is highly sensitive to mismatches between train-time and test-time prompts. Using different prompts for inference than those used during training degraded performance. 

In principle, it's possible to fine-tune only on images and labels, e.g., by training a classifier head or adapting the full model. However, we found that keeping the full prompt (with instruction and query) helps LoRA steer the model effectively, while staying close to the original distribution. This aligns with LoRA’s goal of enabling efficient adaptation without catastrophic forgetting.

As seen in Table~\ref{tab:mirabest_confident}, LoRA fine-tuning struggles with very few labels due to overfitting and limited adaptation of the vision-language representations learned during pretraining. However, once around 145 labels are available, Qwen2 catches up with the ResNet baseline and continues to improve consistently as more data is provided.

%% file: project/text_prompts.tex
\section{Prompts}
\label{app:prompts}
\subsection{Text prompts}

Prompt 0

\begin{nolinenumbers}
\begin{promptlisting}
system_message = (
            "You are an expert radio galaxy classifier. "
            "FR-I: bright core, gradually fading jets. "
            "FR-II: faint core, jets end in hotspots. "
            "Consider core brightness and jet termination."
            )
query_text = 'Respond only FR-I or FR-II.'
\end{promptlisting}
\end{nolinenumbers}

Prompt 1
\begin{nolinenumbers}
\begin{promptlisting}
system_message = 'You are an expert at classifying radio galaxies as either Fanaroff-Riley Type I (FR-I) or Type II (FR-II). FR-I galaxies have central brightness and edge-darkened lobes. FR-II galaxies have edge-brightened lobes and hotspots at lobe ends. Jet characteristics can also aid in classification (common jets for FR-I, often one-sided jets for FR-II). Focus on lobe brightness distribution and hotspot presence.'
query_text = 'Describe the lobe brightness distribution (edge-brightened or edge-darkened) and the presence and location of hotspots in the radio galaxy image. Classify the galaxy as either FR-I or FR-II. Respond with: <answer>FR-I/FR-II</answer>.'
\end{promptlisting}
\end{nolinenumbers}
Prompt 2

\begin{nolinenumbers}
\begin{promptlisting}
system_message='You are an expert at classifying radio galaxies as either Fanaroff-Riley Type I (FR-I) or Type II (FR-II). FR-I galaxies have central brightness and edge-darkened lobes. FR-II galaxies have edge-brightened lobes and hotspots at lobe ends. Jet characteristics can also aid in classification (common jets for FR-I, often one-sided jets for FR-II). Focus on lobe brightness distribution and hotspot presence.'
query_text='Respond only with: <answer>FR-I/FR-II</answer>.'
\end{promptlisting}
\end{nolinenumbers}

Prompt 3

\begin{nolinenumbers}
\begin{promptlisting}
system_message='You are an expert at classifying radio galaxies as either Fanaroff-Riley Type I (FR-I) or Type II (FR-II). FR-I galaxies have central brightness and edge-darkened lobes. FR-II galaxies have edge-brightened lobes and hotspots at lobe ends. Jet characteristics can also aid in classification (common jets for FR-I, often one-sided jets for FR-II). Focus on lobe brightness distribution and hotspot presence.'
query_text='Describe the lobe brightness distribution (edge-brightened or edge-darkened) and the presence and location of hotspots in the radio galaxy image. Classify the galaxy as either FR-I or FR-II. Finish your response with: <answer>FR-I/FR-II</answer>.'
\end{promptlisting}
\end{nolinenumbers}

Prompt 4

\begin{nolinenumbers}
\begin{promptlisting}
system_message='''
You are an astronomer tasked with classifying morphologies of radio galaxies.  
There are two classes:
• FR-I: bright toward the center, fainter at the lobes’ edges (edge-darkened), steep spectra, common jets, ram-pressure distortions in rich X-ray clusters.  
• FR-II: edge-brightened, more luminous, bright hotspots at lobe ends, one-sided jets.'''
query_text='Define a morphology class of this image. Respond only FR-I or FR-II.'
\end{promptlisting}
\end{nolinenumbers}

Prompt 5

\begin{nolinenumbers}
\begin{promptlisting}
system_message='''
You are an astronomer tasked with classifying real images of radio galaxies.
    There are two classes:
    - FR-I: bright toward the center, fainter at the lobes' edges, often shows jets, etc.
    - FR-II: edge-brightened, luminous lobes, bright hotspots at ends.'''
query_text='Define a morphology class of this image. Respond only FR-I or FR-II.'
\end{promptlisting}
\end{nolinenumbers}

Prompt 6

\begin{nolinenumbers}
\begin{promptlisting}
system_message='''
You are an astronomer tasked with classifying morphologies of radio galaxies.  
There are two classes:
• FR-I: bright toward the center, fainter at the lobes’ edges (edge-darkened), common jets, ram-pressure distortions in rich X-ray clusters.  
• FR-II: edge-brightened, more luminous, bright hotspots at lobe ends, one-sided jets.'''
query_text='''Define a morphology class of this image. First analyze the features with respect to the described classes. Conclude if FR-I or FR-II.
Respond to the previous questions in the following format:
<think>...</think>
<answer>...</answer>
'''
\end{promptlisting}
\end{nolinenumbers}

Prompt 7

\begin{nolinenumbers}
\begin{promptlisting}
system_message='''
You are an astronomer tasked with classifying morphologies of radio galaxies.  
There are two classes:
• FR-I: bright toward the center, fainter at the lobes’ edges (edge-darkened), common jets, ram-pressure distortions in rich X-ray clusters.  
• FR-II: edge-brightened, more luminous, bright hotspots at lobe ends, one-sided jets.'''
query_text='''Define a morphology class of this image. First analyze the features with respect to the described classes. Conclude if FR-I or FR-II.
Respond to the previous questions in the following format:
<think>...</think>
<answer>...</answer>
'''
\end{promptlisting}
\end{nolinenumbers}
\newpage
Prompt 8

\begin{nolinenumbers}
\begin{promptlisting}
system_message='''
You are an astronomer tasked with classifying real images of radio galaxies.
    
    There are two classes:
    - FR-I: bright toward the center, fainter at the lobes' edges, often shows jets, etc.
    - FR-II: edge-brightened, luminous lobes, bright hotspots at ends.

    Now you will get the real image of a galaxy. Respond in the following format: <think>...</think> <answer>...</answer>
'''
query_text='''First, list the features you need to identify the class, then analyze them. Estimate the probability from 0 to 1 to be each class. Conclude if FR-I or FR-II.'''
\end{promptlisting}
\end{nolinenumbers}

Prompt 9

\begin{nolinenumbers}
\begin{promptlisting}
system_message='''
You are an astronomer tasked with classifying morphologies of radio galaxies.  
There are two classes:
• FR-I: bright toward the center, fainter at the lobes’ edges (edge-darkened), common jets, ram-pressure distortions in rich X-ray clusters.  
• FR-II: edge-brightened, more luminous, bright hotspots at lobe ends, one-sided jets.
'''
query_text='''Define a morphology class of this image. First analyze the features with respect to the described classes. Conclude which of FR-I or FR-II is more probable (you must choose one class as the answer, you cannot ask for more information or say that you do not know).
Respond to the previous questions in the following format:
    <think>...</think>
    <answer>...</answer>'''
\end{promptlisting}
\end{nolinenumbers}

Prompt 10

\begin{nolinenumbers}
\begin{promptlisting}
system_message=(
            "Core: Bright or Faint? Jets: Fading or Hotspots? "
            "<think>Classify based on core and jet properties: FR-I (bright core, fading jets) "
            "or FR-II (faint core, hotspots). Make a selection.</think> "
            "<answer>FR-I or FR-II?</answer>"
)
query_text=(
            "Respond only"
            "<answer>FR-I or FR-II?</answer>"
        )
\end{promptlisting}
\end{nolinenumbers}
Prompt 11

\begin{nolinenumbers}
\begin{promptlisting}
system_message='You are an expert at classifying radio galaxies as either Fanaroff-Riley Type I (FR-I) or Type II (FR-II). FR-I galaxies have central brightness and edge-darkened lobes. FR-II galaxies have edge-brightened lobes and hotspots at lobe ends. Jet characteristics can also aid in classification (common jets for FR-I, often one-sided jets for FR-II). Focus on lobe brightness distribution and hotspot presence.'
query_text='Describe the lobe brightness distribution (edge-brightened or edge-darkened) and the presence and location of hotspots in the radio galaxy image. Classify the galaxy as either FR-I or FR-II. Finish your response with: <answer>FR-I/FR-II</answer>.'
\end{promptlisting}
\end{nolinenumbers}

%% file: project/diagram_prompts.tex
\subsection{Diagram}

Prompt 0

\begin{nolinenumbers}
\begin{promptlisting}
system_message='''
You are an astronomer tasked with classifying real images of radio galaxies based on this diagram.
    The image above is a diagram comparing two radio galaxy morphologies. Ignore it for classification, remember just features.
    
    There are two classes:
    - FR-I: bright toward the center, fainter at the lobes' edges, often shows jets, etc.
    - FR-II: edge-brightened, luminous lobes, bright hotspots at ends.'''
query_text='Define a morphology class of this image. Respond only FR-I or FR-II.'
\end{promptlisting}
\end{nolinenumbers}

Prompt 1

\begin{nolinenumbers}
\begin{promptlisting}
system_message='''
You are an astronomer tasked with classifying real images of radio galaxies.
    There are two classes:
    - FR-I: bright toward the center, fainter at the lobes' edges, often shows jets, etc.
    - FR-II: edge-brightened, luminous lobes, bright hotspots at ends.'''
query_text='Define a morphology class of this image. Respond only FR-I or FR-II.'
\end{promptlisting}
\end{nolinenumbers}

Prompt 2

\begin{nolinenumbers}
\begin{promptlisting}
system_message='''You are an astronomer tasked with classifying real images of radio galaxies based on the diagram comparing two radio galaxy morphologies based on their features. Ignore it for classification, remember just features.
    
    There are two classes:
    - FR-I: bright toward the center, fainter at the lobes' edges, often shows jets, etc.
    - FR-II: edge-brightened, luminous lobes, bright hotspots at ends.
    
    When studying the query image, analyze each listed feature separately.'''
query_text='''Define a morphology class of this image. 
    Analyze the features with respect to the described classes. 
    Conclude if FR-I or FR-II.
    Respond to the previous questions in the following format: <think>...</think> <answer>...</answer>'''
\end{promptlisting}
\end{nolinenumbers}

Prompt 3

\begin{nolinenumbers}
\begin{promptlisting}
system_message='''You are an astronomer tasked with classifying real images of radio galaxies.
    This diagram roughly shows the features of two classes:
    - FR-I: bright toward the center, fainter at the lobes' edges, often shows jets, etc.
    - FR-II: edge-brightened, luminous lobes, bright hotspots at ends.
    '''
query_text='''Define a morphology class of this image. First analyze the features with respect to the described classes. Conclude if FR-I or FR-II.
Respond to the previous questions in the following format:
<think>...</think>
<answer>...</answer>'''
\end{promptlisting}
\end{nolinenumbers}

Prompt 4

\begin{nolinenumbers}
\begin{promptlisting}
system_message='''You are an astronomer tasked with classifying real images of radio galaxies.
    
    There are two classes:
    - FR-I: bright toward the center, fainter at the lobes' edges, often shows jets, etc.
    - FR-II: edge-brightened, luminous lobes, bright hotspots at ends.

    Now you will get the real image of a galaxy. Respond in the following format: <think>...</think> <answer>...</answer>'''
query_text='''List the features you need to identify the class. Conclude if FR-I or FR-II.'''
\end{promptlisting}
\end{nolinenumbers}

Prompt 5

\begin{nolinenumbers}
\begin{promptlisting}
system_message='''
You are an astronomer tasked with classifying real images of radio galaxies.
    
    There are two classes:
    - FR-I: bright toward the center, fainter at the lobes' edges, often shows jets, etc.
    - FR-II: edge-brightened, luminous lobes, bright hotspots at ends.

    Now you will get the real image of a galaxy. Respond in the format: <think>...</think> <answer>...</answer>. '''
query_text='''First, list the features you need to identify the class, then analyze them. Estimate the probability from 0 to 1 to be each class. Conclude if FR-I or FR-II.'''
\end{promptlisting}
\end{nolinenumbers}

Prompt 6

\begin{nolinenumbers}
\begin{promptlisting}
system_message='''
You are an astronomer tasked with classifying real images of radio galaxies based on the diagram comparing two radio galaxy morphologies based on their features. Ignore it for classification, remember just features.
    
    There are two classes:
    - FR-I: bright toward the center, fainter at the lobes' edges, often shows jets, etc.
    - FR-II: edge-brightened, luminous lobes, bright hotspots at ends.
    
    When studying the query image, analyze each listed feature separately.'''
query_text='''First, list the features you need to identify the class, then analyze them. Carefully consider each feature.Keep the reasoning short. After always classify the last image into one of the two classes. Conclude your final answer as:
<answer>FR-I</answer>
or
<answer>FR-II</answer>'''
\end{promptlisting}
\end{nolinenumbers}

Prompt 7

\begin{nolinenumbers}
\begin{promptlisting}
system_message = """You are an astronomer tasked with classifying real images of radio galaxies based on this diagram.
    The image above is a diagram comparing two radio galaxy morphologies.
    
    There are two classes:
    - FR-I: bright toward the center, fainter at the lobes' edges, often shows jets, etc.
    - FR-II: edge-brightened, luminous lobes, bright hotspots at ends.
    
    Respond with only FR-I or FR-II."""
query_text=''
\end{promptlisting}
\end{nolinenumbers}

%% file: project/few_shot_prompt.tex
\subsection{Few-shot prompts}

Prompt 0

\begin{nolinenumbers}
\begin{promptlisting}
system_message='''
You are an astronomer tasked with classifying morphologies of radio galaxies.  
There are two classes:
• FR-I: bright toward the center, fainter at the lobes’ edges (edge-darkened), steep spectra, common jets, ram-pressure distortions in rich X-ray clusters.  
• FR-II: edge-brightened, more luminous, bright hotspots at lobe ends, one-sided jets.'''
query_text='Define a morphology class of this image. Respond only FR-I or FR-II.'
examples_message=""
query_text_example='Define a morphology class of this image. Respond only FR-I or FR-II.'
\end{promptlisting}
\end{nolinenumbers}

Prompt 1

\begin{nolinenumbers}
\begin{promptlisting}
system_message='''
You are an astronomer tasked with classifying real images of radio galaxies.
    There are two classes:
    - FR-I: bright toward the center, fainter at the lobes' edges, often shows jets, etc.
    - FR-II: edge-brightened, luminous lobes, bright hotspots at ends.
   
'''
query_text='Define a morphology class of this image. Respond only FR-I or FR-II.'
examples_message="Here are examples:"
query_text_example='Define a morphology class of this image. Respond only FR-I or FR-II.'
\end{promptlisting}
\end{nolinenumbers}

Prompt 2

\begin{nolinenumbers}
\begin{promptlisting}
system_message='''

 You are an astronomer classifying the morphology of real galaxies. 
            
    There are two classes:
    - FR-I: bright toward the center, fainter at the lobes' edges, often shows jets, etc.
    - FR-II: edge-brightened, luminous lobes, bright hotspots at ends.

    Look carefully through examples of radio galaxies:'''
query_text='''
    Explain the previous examples. Then classify the following galaxy image using the format <think>..</think><answer>..</answer>.  Explain your reasoning and motivation. '''
examples_message=""
query_text_example='Example: Classify this galaxy image.'
\end{promptlisting}
\end{nolinenumbers}

Prompt 3

\begin{nolinenumbers}
\begin{promptlisting}
system_message='''You are an astronomer classifying the morphology of real galaxies. 
            
    There are two classes:
    - FR-I: bright toward the center, fainter at the lobes' edges, often shows jets, etc.
    - FR-II: edge-brightened, luminous lobes, bright hotspots at ends.

    Format your answer as <think></think><answer></answer>. 
 Classify the galaxy image. '''
query_text=""
examples_message=""
query_text_example='Example: Classify this galaxy image.'
\end{promptlisting}
\end{nolinenumbers}

Prompt 4

\begin{nolinenumbers}
\begin{promptlisting}
system_message='''You are an astronomer classifying the morphology of real galaxies. 
            
    There are two classes:
    - FR-I: bright toward the center, fainter at the lobes' edges, often shows jets, etc.
    - FR-II: edge-brightened, luminous lobes, bright hotspots at ends.

    Format your answer as <think></think><answer></answer>. 
 Classify the galaxy image. '''
query_text='''Explain how to correctly classify radio galaxies as FR-I or FR-II. Classify the galaxy image. '''
examples_message=""
query_text_example='Classify this galaxy image.'
\end{promptlisting}
\end{nolinenumbers}

Prompt 5

\begin{nolinenumbers}
\begin{promptlisting}
system_message='''You are an astronomer classifying the morphology of real galaxies. 
            
    There are two classes:
    - FR-I: bright toward the center, fainter at the lobes' edges, often shows jets, etc.
    - FR-II: edge-brightened, luminous lobes, bright hotspots at ends.

    Format your answer as <think></think><answer></answer>. 
 Classify the galaxy image. '''
query_text='''Classify the last image. Use the format <think>..</think><answer>..</answer>.  In your think block explain thoroughly your class prediction.'''
examples_message=""
query_text_example='Classify this galaxy image.'
\end{promptlisting}
\end{nolinenumbers}

Prompt 6

\begin{nolinenumbers}
\begin{promptlisting}
system_message = (
            "You are an expert radio galaxy classifier. "
            "FR-I: bright core, gradually fading jets. "
            "FR-II: faint core, jets end in hotspots. "
            "Consider core brightness and jet termination."
            )
query_text = (
            "Core: Bright or Faint? Jets: Fading or Hotspots? "
            "Classify based on core and jet properties: FR-I (bright core, fading jets) "
            "or FR-II (faint core, hotspots). Make a selection."
            "FR-I or FR-II?"
        )
query_text_example = "Respond only FR-I or FR-II."
examples_message = ""
summary_after_examples_text = ""
\end{promptlisting}
\end{nolinenumbers}

Prompt 7

\begin{nolinenumbers}
\begin{promptlisting}
system_message='''You are an astronomer tasked with classifying real images of radio galaxies.
    There are two classes:
    - FR-I: bright toward the center, fainter at the lobes' edges, often shows jets, etc.
    - FR-II: edge-brightened, luminous lobes, bright hotspots at ends.
    
    '''
query_text='Define a morphology class of this image. Respond only FR-I or FR-II.'
examples_message=""
query_text_example='Define a morphology class of this image. Respond only FR-I or FR-II.'
\end{promptlisting}
\end{nolinenumbers}

%% file: project/extra_tables.tex
\section{Full results tables}

\begin{table}[ht]
\centering
\scriptsize
\caption{Text \;|\; image after the query \;|\; Error (\%)  }
\begin{tabular}{rrrrrrr}
\toprule
\textbf{index} & \textbf{LLaVA-1.6} & \textbf{Qwen2-2B} & \textbf{Qwen2.5-3B} & \textbf{Qwen2-7B} & \textbf{Qwen2.5-7B} & \textbf{Gemini}\\
\midrule
0  & \textit{42.3}   & 52.9   & 60.6   & 52.9   & 49.0   & \textbf{19.2}   \\
1  & 47.1   & 50.0   & \textit{46.2}   & 51.9   & 52.9   & \textbf{26.9}   \\
2  & 47.1   & 44.2   & 52.9   & \textit{43.3}   & 58.7   & \textbf{25.0}   \\
3  & 47.1   & 56.7   & 64.4   & \textit{40.4}   & 51.0   & \textbf{31.7}   \\
4  & 48.1   & 47.1   & 45.2   & \textit{43.3}   & 50.0   & \textbf{23.1}   \\
5  & 54.8   & 43.3   & 53.8   & \textit{34.6}   & 43.3   & \textbf{17.3}   \\
6  & 50.0   & 53.8   & \textit{38.5}   & 40.4   & 45.2   & \textbf{29.8}   \\
7  & 50.0   & 53.8   & \textit{37.5}   & 40.4   & 42.3   & \textbf{30.8}   \\
8  & 47.1   & 40.4   & 39.4   & \textit{33.7}   & 42.3   & \textbf{13.5}   \\
9 & 53.8   & \textit{43.3}   & 50.0   & \textit{43.3}   & 50.0   & \textbf{33.7}   \\
10 & 47.1   & 52.9   & \textit{46.2}   & 48.1   & 52.9   & \textbf{28.8}   \\
11 & 47.1   & 56.7   & 55.8   & \textit{40.4}   & 56.7   & \textbf{30.8}   \\
\bottomrule
\end{tabular}
\label{tab:text_0}
\end{table}

\begin{table}[ht]
\centering
\scriptsize
\caption{Diagram \;|\; image after the query \;|\; Error (\%)  }
\begin{tabular}{rrrrrrr}
\toprule
\textbf{index} & \textbf{LLaVA-1.6} & \textbf{Qwen2-2B} & \textbf{Qwen2.5-3B} & \textbf{Qwen2-7B} & \textbf{Qwen2.5-7B} & \textbf{Gemini}\\
\midrule
0 & 62.5   & 47.1   & 50.0   & 46.2   & \textit{46.2}   & \textbf{19.2}   \\
1 & 57.7   & 47.1   & 47.1   & \textit{37.5}   & 46.2   & \textbf{16.3}   \\
2 & 49.0   & 55.8   & 43.3   & \textit{40.4}   & 45.2   & \textbf{29.8}   \\
3 & 47.1   & 43.3   & 48.1   & \textit{42.3}   & 51.0   & \textbf{32.7}   \\
4 & \textit{35.6}   & 47.1   & 51.9   & 36.5   & 49.0   & \textbf{21.2}   \\
5 & 45.2   & 47.1   & 48.1   & \textit{28.8}   & 48.1   & \textbf{20.2}   \\
6 & 54.8   & 53.8   & 55.8   & \textit{44.2}   & 49.0   & \textbf{18.3}   \\
7 & 46.2   & 48.1   & 46.2   & \textit{36.5}   & 39.4   & \textbf{16.3}   \\
\bottomrule
\end{tabular}
\label{tab:diagram_0}
\end{table}

\begin{table}[ht]
\centering
\scriptsize
\caption{Fixed-Imgs \;|\; image after the query \;|\; Error (\%)  }
\begin{tabular}{rrrrrrr}
\toprule
\textbf{index} & \textbf{LLaVA-1.6} & \textbf{Qwen2-2B} & \textbf{Qwen2.5-3B} & \textbf{Qwen2-7B} & \textbf{Qwen2.5-7B} & \textbf{Gemini}\\
\midrule
0 & 47.1   & 44.2   & 26.9   & \textbf{22.1}   & 46.2   & \textit{26.9}   \\
1 & 47.1   & 36.5   & 28.8   & \textit{24.0}   & 32.7   & \textbf{18.3}   \\
2 & 58.7   & \textbf{29.8}   & 46.2   & 37.5   & 41.3   & \textit{30.8}   \\
3 & 76.9   & \textit{27.9}   & 34.6   & \textbf{19.2}   & 29.8   & 51.9   \\
4 & 65.4   & 33.7   & 49.0   & \textit{24.0}   & 42.3   & \textbf{19.2}   \\
5 & 100.0   & \textit{28.8}   & 50.0   & 35.6   & 30.8   & \textbf{16.3}   \\
6 & 52.9   & 45.2   & 46.2   & \textit{29.8}   & 51.0   & \textbf{21.2}   \\
7 & 47.1   & 39.4   & 31.7   & \textit{20.2}   & 31.7   & \textbf{19.2}   \\
\bottomrule
\end{tabular}
\label{tab:fixed_imgs_0}
\end{table}

\begin{table}[ht]
\centering
\scriptsize
\caption{kNN-Imgs \;|\; image after the query \;|\; Error (\%)  }
\begin{tabular}{rrrrrrr}
\toprule
\textbf{index} & \textbf{LLaVA-1.6} & \textbf{Qwen2-2B} & \textbf{Qwen2.5-3B} & \textbf{Qwen2-7B} & \textbf{Qwen2.5-7B} & \textbf{Gemini}\\
\midrule
0 & 29.8   & 37.5   & 30.8   & \textit{26.0}   & 28.8   & \textbf{20.2}   \\
1 & 28.8   & 35.6   & 27.9   & \textbf{16.3}   & 24.0   & \textit{18.3}   \\
2 & 38.5   & \textbf{26.0}   & 38.5   & \textit{26.9}   & 29.8   & 34.6   \\
3 & 60.6   & \textbf{26.0}   & 40.4   & \textit{26.9}   & 32.7   & 31.7   \\
4 & 32.7   & 25.0   & 46.2   & \textit{23.1}   & 39.4   & \textbf{19.2}   \\
5 & 52.9   & 27.9   & 34.6   & \textit{23.1}   & 31.7   & \textbf{22.1}   \\
6 & 25.0   & 29.8   & 33.7   & \textit{24.0}   & 28.8   & \textbf{9.6}   \\
7 & 28.8   & 38.5   & 29.8   & \textbf{16.3}   & 24.0   & \textit{21.2}   \\
\bottomrule
\end{tabular}
\label{tab:knn_imgs_0}
\end{table}

\begin{table}[ht]
\centering
\scriptsize
\caption{kNN-Balanced \;|\; image after the query \;|\; Error (\%)  }
\begin{tabular}{rrrrrrr}
\toprule
\textbf{index} & \textbf{LLaVA-1.6} & \textbf{Qwen2-2B} & \textbf{Qwen2.5-3B} & \textbf{Qwen2-7B} & \textbf{Qwen2.5-7B} & \textbf{Gemini}\\
\midrule
0 & 47.1   & 50.0   & 45.2   & \textit{37.5}   & 27.9   & \textbf{26.0}   \\
1 & 47.1   & 35.6   & 41.3   & 32.7   & \textit{25.0}   & \textbf{18.3}   \\
2 & \textit{48.1}   & \textbf{41.3}   & 51.0   & \textbf{41.3}   & 51.0   & 68.3   \\
3 & 62.5   & 40.4   & \textbf{33.7}   & \textit{34.6}   & 36.5   & 45.2   \\
4 & 52.9   & 46.2   & \textit{39.4}   & 50.0   & 47.1   & \textbf{26.9}   \\
5 & 80.8   & \textit{47.1}   & 49.0   & 55.8   & \textit{47.1}   & \textbf{11.5}   \\
6 & 51.9   & 47.1   & 40.4   & \textit{39.4}   & 51.9   & \textbf{19.2}   \\
7 & 47.1   & 34.6   & 39.4   & \textit{31.7}   & 30.8   & \textbf{19.2}   \\
\bottomrule
\end{tabular}
\label{tab:knn_balanced_0}
\end{table}

\begin{table}[ht]
\centering
\scriptsize
\caption{Text \;|\; image after the query \;|\; Error (\%)}
\begin{tabular}{rrrrrrr}
\toprule
\textbf{index} & \textbf{LLaVA-1.6} & \textbf{Qwen2-2B} & \textbf{Qwen2.5-3B} & \textbf{Qwen2-7B} & \textbf{Qwen2.5-7B} & \textbf{Gemini}\\
\midrule
0  & \textit{42.3}   & 52.9   & 63.5   & 49.0   & 44.2   & \textbf{22.1}   \\
1  & 47.1   & 51.9   & 47.1   & \textit{38.5}   & 45.2   & \textbf{24.0}   \\
2  & \textit{47.1}   & 52.9   & 51.0   & 48.1   & 51.0   & \textbf{28.8}   \\
3  & 47.1   & 50.0   & 47.1   & \textit{39.4}   & 51.9   & \textbf{26.0}   \\
4  & 48.1   & 43.3   & 51.9   & 52.9   & \textit{42.3}   & \textbf{23.1}   \\
5  & 54.8   & 49.0   & 43.3   & 46.2   & \textit{31.7}   & \textbf{25.0}   \\
6  & 50.0   & 51.9   & 47.1   & \textbf{34.6}   & 43.3   & \textit{38.5}   \\
7  & 50.0   & 51.9   & 47.1   & \textbf{34.6}   & \textit{44.2}   & \textbf{34.6}   \\
8  & 47.1   & 98.1   & 44.2   & 43.3   & \textit{38.5}   & \textbf{26.0}   \\
9 & 53.8   & 49.0   & 58.7   & \textit{40.4}   & 49.0   & \textbf{29.8}   \\
10 & \textit{47.1}   & 52.9   & 52.9   & 52.9   & 52.9   & \textbf{26.0}   \\
11 & 47.1   & 50.0   & 51.0   & \textit{39.4}   & 50.0   & \textbf{28.8}   \\
\bottomrule
\end{tabular}
\label{tab:text_1}
\end{table}

\begin{table}[ht]
\centering
\scriptsize
\caption{Diagram \;|\; image after the query \;|\; Error (\%)  }
\begin{tabular}{rrrrrrr}
\toprule
\textbf{index} & \textbf{LLaVA-1.6} & \textbf{Qwen2-2B} & \textbf{Qwen2.5-3B} & \textbf{Qwen2-7B} & \textbf{Qwen2.5-7B} & \textbf{Gemini}\\
\midrule
0 & 62.5   & \textit{47.1}   & 50.0   & 51.9   & \textit{47.1}   & \textbf{17.3}   \\
1 & 57.7   & \textit{47.1}   & 50.0   & 52.9   & \textit{47.1}   & \textbf{18.3}   \\
2 & 49.0   & 46.2   & 46.2   & 49.0   & \textit{40.4}   & \textbf{37.5}   \\
3 & 47.1   & 49.0   & 46.2   & \textit{32.7}   & 49.0   & \textbf{31.7}   \\
4 & \textit{35.6}   & 51.9   & 43.3   & 41.3   & 46.2   & \textbf{23.1}   \\
5 & \textit{45.2}   & 51.9   & 50.0   & 49.0   & 47.1   & \textbf{25.0}   \\
6 & 54.8   & 52.9   & \textit{48.1}    & \textit{48.1}   & 56.7   & \textbf{15.4}   \\
7 & 46.2   & 48.1   & 48.1   & \textit{36.5}   & 41.3   & \textbf{22.1}   \\
\bottomrule
\end{tabular}
\label{tab:diagram_1}
\end{table}

\begin{table}[ht]
\centering
\scriptsize
\caption{Fixed-Imgs \;|\; image after the query \;|\; Error (\%)  }
\begin{tabular}{rrrrrrr}
\toprule
\textbf{index} & \textbf{LLaVA-1.6} & \textbf{Qwen2-2B} & \textbf{Qwen2.5-3B} & \textbf{Qwen2-7B} & \textbf{Qwen2.5-7B} & \textbf{Gemini}\\
\midrule
0 & 47.1   & 43.3   & \textit{27.9}   & 46.2   & 42.3   & \textbf{19.2}   \\
1 & 47.1   & 36.5   & 26.0   & \textit{24.0}   & 30.8   & \textbf{18.3}   \\
2 & 58.7   & \textbf{37.5}   & \textit{50.0}   & 60.6   & 55.8   & 57.7   \\
3 & 76.9   & \textit{23.1}   & 35.6   & \textbf{18.3}   & 27.9   & 52.9   \\
4 & 65.4   & 52.9   & 48.1   & \textit{47.1}   & 42.3   & \textbf{22.1}   \\
5 & 100.0   & 44.2   & 51.9   & \textit{36.5}   & 51.0   & \textbf{24.0}   \\
6 & 52.9   & 47.1   & \textit{28.8}   & 32.7   & 51.0   & \textbf{21.2}   \\
7 & 47.1   & 41.3   & \textit{26.9}   & \textit{26.9}   & 29.8   & \textbf{21.2}   \\
\bottomrule
\end{tabular}
\label{tab:fixed_imgs_1}
\end{table}

\begin{table}[ht]
\centering
\scriptsize
\caption{kNN-Imgs \;|\; image\_first=1 \;|\; Error (\%)  }
\begin{tabular}{rrrrrrr}
\toprule
\textbf{index} & \textbf{LLaVA-1.6} & \textbf{Qwen2-2B} & \textbf{Qwen2.5-3B} & \textbf{Qwen2-7B} & \textbf{Qwen2.5-7B} & \textbf{Gemini}\\
\midrule
0 & 29.8   & 40.4   & \textit{27.9}   & 29.8   & 29.8   & \textbf{16.3}   \\
1 & 28.8   & 40.4   & 31.7   & \textit{23.1}   & 27.9   & \textbf{19.2}   \\
2 & \textit{38.5}   & 46.2   & \textbf{33.7}   & 41.3   & 39.4   & 80.8   \\
3 & 60.6   & 30.8   & 29.8   & \textbf{24.0}   & 29.8   & \textit{28.8}   \\
4 & 32.7   & 29.8   & 32.7   & \textit{26.9}   & 32.7   & \textbf{24.0}   \\
5 & 52.9   & 31.7   & 32.7   & 29.8   & \textit{27.9}   & \textbf{17.3}   \\
6 & 25.0   & 31.7   & 24.0   & \textit{23.1}   & 27.9   & \textbf{17.3}   \\
7 & 28.8   & 51.0   & 27.9   & \textit{24.0}   & 26.0   & \textbf{14.4}   \\
\bottomrule
\end{tabular}
\label{tab:knn_imgs_1}
\end{table}

\begin{table}[ht]
\centering
\scriptsize
\caption{kNN-Balanced \;|\; image\_first=1 \;|\; Error (\%)  }
\begin{tabular}{rrrrrrr}
\toprule
\textbf{index} & \textbf{LLaVA-1.6} & \textbf{Qwen2-2B} & \textbf{Qwen2.5-3B} & \textbf{Qwen2-7B} & \textbf{Qwen2.5-7B} & \textbf{Gemini}\\
\midrule
0 & 47.1   & 50.0   & 45.2   & 37.5   & \textit{27.9}   & \textbf{26.0}   \\
1 & 47.1   & 35.6   & 41.3   & 32.7   & \textit{25.0}   & \textbf{18.3}   \\
2 & \textit{48.1}   & \textbf{41.3}   & 51.0   & \textbf{41.3}   & 51.0   & 68.3   \\
3 & 62.5   & 40.4   & \textbf{33.7}   & \textit{34.6}   & 36.5   & 45.2   \\
4 & 52.9   & 46.2   & \textit{39.4}   & 50.0   & 47.1   & \textbf{26.9}   \\
5 & 80.8   & \textit{47.1}   & 49.0   & 55.8   & \textit{47.1}   & \textbf{11.5}   \\
6 & 51.9   & 47.1   & 40.4   & \textit{39.4}   & 51.9   & \textbf{19.2}   \\
7 & 47.1   & 34.6   & 39.4   & 31.7   & \textit{30.8}   & \textbf{19.2}   \\
\bottomrule
\end{tabular}
\label{tab:knn_balanced_1}
\end{table}